\documentclass[aps,prx,reprint,superscriptaddress,amsmath,amssymb]{revtex4-2}

\usepackage{graphicx}
\usepackage{txfonts}
\usepackage[hypertexnames=false]{hyperref}
\usepackage{braket}
\usepackage{bm}
\usepackage{enumitem}

\begin{document}

\title{Bubble formation in active binary mixture model}

\author{Kyosuke Adachi}
\affiliation{RIKEN Center for Interdisciplinary Theoretical and Mathematical Sciences, 2-1 Hirosawa, Wako 351-0198, Japan}
\affiliation{Nonequilibrium Physics of Living Matter Laboratory, RIKEN Pioneering Research Institute, 2-1 Hirosawa, Wako 351-0198, Japan}

\date{\today}

\begin{abstract}
Phase separation, the spontaneous segregation of density, is a ubiquitous phenomenon observed across diverse physical and biological systems.
Within a crowd of motile elements, active phase separation emerges from the interplay of activity (i.e., self-propulsion) and density interactions.
A striking feature of active phase separation is the persistent formation of dilute-phase bubbles within the dense phase, which has been explored in theoretical models.
However, the fundamental parameters that systematically control bubble formation remain unclear in conventional self-propelled particle models.
Here, we introduce an active binary mixture model in which solutes and solvents dynamically exchange positions on a lattice; both solutes and solvents are self-propelled particles, but solvents play a role analogous to empty space in typical dry active matter.
Within this model, we find that spontaneous bubble formation of solvents can be tuned by activity asymmetry, which is the difference between the solute and solvent activities.
Numerical simulations reveal that moderate solvent activity enhances bubble formation, while larger solvent activity, comparable to solute activity, suppresses it.
By employing mean-field theory, which captures essential phase behaviors, we consider the mechanism for the enhancement of bubble formation induced by solvent activity.
Beyond these findings, when solute and solvent activities are equal, we apply the finite-size scaling analysis to estimate the critical exponents for active phase separation under the suppression of bubbles.
Our findings establish activity asymmetry as a key control parameter for active matter phase transitions, offering new insights into universality in nonequilibrium systems.
\end{abstract}

\maketitle

\section{Introduction}

Active matter exhibits diverse collective phenomena, ranging from giant number fluctuations~\cite{Ramaswamy2003, Shankar2018} to active turbulence~\cite{Dunkel2013}, depending on the type of activity of constituents and their interactions~\cite{Marchetti2013, Gompper2025, Te-Vrugt2025}.
Experimentally, collective properties of active matter have been investigated extensively using biological matter~\cite{Szabo2006, Nishiguchi2017, Kawaguchi2017, Liu2019, Tan2022} and artificial systems~\cite{Deseigne2010, Palacci2013, Bricard2013, Buttinoni2013, Kumar2014, Ginot2018, Deblais2018, Geyer2019, Chardac2021, Wang2021}.
From theoretical viewpoints, the effects of activity have been studied using particle models such as the Vicsek model~\cite{Vicsek1995} and coarse-grained models such as the Toner-Tu model~\cite{Toner1995}.
Activity has been shown to produce material phases that are distinct from equilibrium counterparts~\cite{Needleman2017, Chate2020}; for example, flocking transitions can spontaneously break continuous symmetry even in two dimensions~\cite{Toner1995, Toner1998, Toner2012, Toner2012mtt, Mahault2019} or lead to metastable ordering~\cite{Codina2022, Besse2022, Benvegnen2023}.

Active phase separation~\cite{Cates2024}, also called motility-induced phase separation~\cite{Cates2015}, is a prototypical example of collective phenomena in active matter, where activity and density-dependent interactions induce the self-aggregation of particles.
Detailed properties of active phase separation have been investigated using particle models such as active Brownian particles~\cite{Fily2012}, active Ornstein-Uhlenbeck particles~\cite{Martin2021}, and active lattice gas~\cite{Thompson2011}, as well as coarse-grained partial differential equation models called Active Model B (AMB)~\cite{Wittkowski2014} and Active Model B+ (AMB+)~\cite{Tjhung2018}.
Since the microscopic mechanism for active phase separation is distinct from that for equilibrium phase separation, special features of the former, such as spontaneous velocity alignment~\cite{Caprini2020}, have been found in numerical studies.
However, the emergence of their similarities in macroscale behaviors has also been discussed~\cite{Tailleur2008, Cates2015}; for example, effective free energy has been constructed for AMB~\cite{Solon2018, Solon2018full}.

Compared to equilibrium phase separation, persistent formation of dilute-phase bubbles within the dense phase is a unique feature of active phase separation~\cite{Tjhung2018, Caballero2018, Singh2019, Shi2020, Caporusso2020, Nakano2024, Fausti2024}.
In particular, AMB+ includes two kinds of nonequilibrium terms that reflect effects of activity: the $\lambda$ term, which defines a local chemical potential, and the $\zeta$ term, which controls a nonlocal chemical potential~\cite{Tjhung2018}.
A mean-field analysis of AMB+ has suggested that the $\lambda$ and $\zeta$ terms, which break time-reversal symmetry and the symmetry between the dense and dilute phases, are crucial for bubble formation~\cite{Tjhung2018}.
Bubbles qualitatively change the phase behavior in large systems, leading to bubbly phase separation~\cite{Tjhung2018, Fausti2024} or microphase separation~\cite{Tjhung2018, Shi2020} instead of macroscopic phase separation, which is typically observed in equilibrium~\cite{Rubinstein2003}.
Despite the significance of bubbles in active phase separation, it is still unclear what parameters can systematically control bubble formation in active particle models, except for a few numerical examples~\cite{Caporusso2020}.
In addition, large-scale and long-time simulations are required to investigate the properties of bubbles in typical particle models~\cite{Shi2020}.

\begin{figure*}[t]
    \centering
    \includegraphics[scale=1]{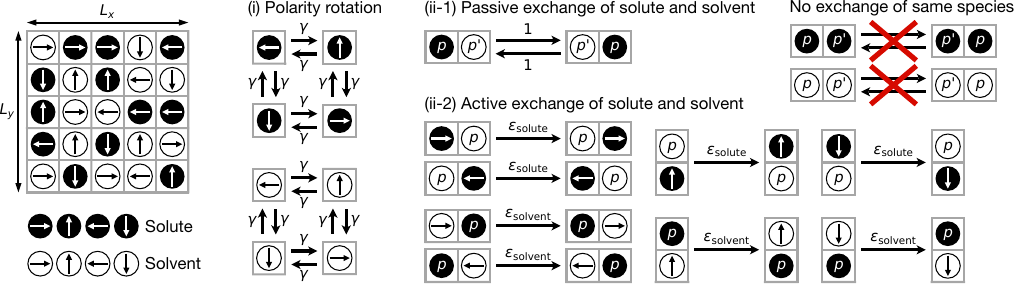}
    \caption{Active binary mixture model.
    Each site of a rectangular lattice with size $(L_x, L_y)$ is occupied by a solute with polarity (black particle with arrow) or a solvent with polarity (white particle with arrow).
    Each solute (solvent) stochastically (i) rotates its polarity by $\pm 90^\circ$ or (ii) exchanges its position with a neighboring solvent (solute).
    No exchange between two solutes or between two solvents is allowed.
    The polarity rotation rate is $\gamma$ [see (i)], the passive exchange rate is $1$ [see (ii-1)], and the active exchange rate for solute (solvent) is $\varepsilon_\mathrm{solute}$ ($\varepsilon_\mathrm{solvent}$) [see (ii-2)].}
    \label{fig_model}
\end{figure*}

Bubbles have also been suspected to affect the critical behavior of active phase separation~\cite{Caballero2018, Shi2020, Speck2022, Cates2024}, which may~\cite{Partridge2019, Maggi2021, Gnan2022} or may not~\cite{Siebert2018, Dittrich2021} belong to the Ising universality class, i.e., the universality class for equilibrium phase separation~\cite{Kadanoff1967}, potentially depending on dimensionality~\cite{Omar2021prl, Feng2025}.
To address this question using active particle models, investigation of critical phenomena under the suppression of bubbles would provide a useful baseline; in the regime with negligible effects of bubbles, the Ising universality is expected to appear, according to renormalization group analyses of coarse-grained models~\cite{Caballero2018, Partridge2019}.
More generally, by highlighting critical phenomena, active matter systems can be systematically connected to nonequilibrium models in other disciplines, such as socioeconomic systems~\cite{Zakine2024}.

In this study, we propose an active binary mixture model, where active solutes and solvents stochastically exchange their positions on a lattice.
By conducting numerical simulations, we find that bubble formation in active phase separation is controlled by the asymmetry in activity between solutes and solvents.
Without solvent activity, this model is equivalent to active lattice gas and undergoes phase separation by solute activity, where bubbles of solvents can gradually appear as the system size increases.
In contrast, with moderate solvent activity, we observe enhanced bubble formation even in relatively small systems, leading to a power-law distribution of the bubble size.
Mean-field theory qualitatively reproduces the nonstationary state with persistent bubble formation, even though the mean-field dynamics is deterministic.
The observed polarity density of solvents suggests that active solvents help bubbles expand within the solute-rich phase.
Larger solvent activity that is comparable to solute activity, in turn, suppresses bubbles.
This is presumably due to the macroscopic symmetry between the solute-rich and solvent-rich phases; in this case, there is no reason for bubbles to grow within either phase.
Near the phase-separation transition line, we find that solute and solvent activities compete with each other, which is explained by linear stability analysis of mean-field theory.
Lastly, when solute and solvent activities are equal, we estimate the critical exponents for active phase separation under the suppression of bubbles, though we need larger-scale simulations to determine the universality class.

\begin{figure*}[t]
    \centering
    \includegraphics[scale=1]{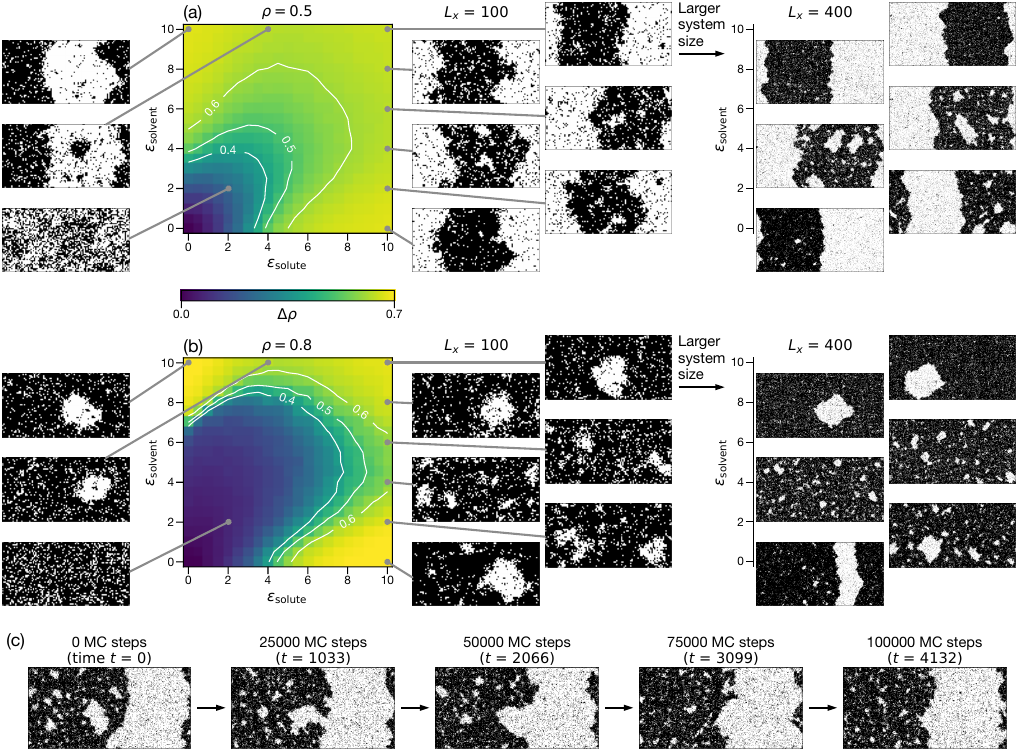}
    \caption{Qualitative phase diagram for active phase separation.
    (a, b) Heatmap of $\Delta \rho$, the shifted difference in solute density between the dense and dilute regions, for $(L_x, L_y) = (100, 50)$ with (a) $\rho = 0.5$ or (b) $\rho = 0.8$.
    The brighter color indicates a higher degree of phase separation, and the white lines represent constant-$\Delta \rho$ lines.
    We show typical snapshots at several sets of activities $(\varepsilon_\mathrm{solute}, \varepsilon_\mathrm{solvent})$, where black and white colors indicate solutes and solvents, respectively.
    We also present several snapshots for a larger system size [$(L_x, L_y) = (400, 200)$].
    (c) Typical time evolution of bubbles in the steady state for $(\varepsilon_\mathrm{solute}, \varepsilon_\mathrm{solvent}) = (10, 4)$.
    Bubbles form and grow within the bulk solute-rich phase and merge into the bulk solvent-rich phase.
    We conducted Monte Carlo (MC) simulations of the active binary mixture model (Fig.~\ref{fig_model}) by sequentially updating the configuration according to the probabilities of rotation and exchange based on time discretization with interval $\Delta t$ for a single MC step.
    We used $\Delta t \simeq 0.0413$ for the case of (c), where time is measured in units of the inverse of the passive exchange rate [see Eq.~\eqref{appeq_timestep} in Appendix~\ref{app_implementation_of_stochastic_dynamics} for the general parameter dependence of $\Delta t$].}
    \label{fig_phase_diagram}
\end{figure*}

\section{Active binary mixture model}
\label{sec_lattice_model}

Mixtures of active particles with distinct activities (i.e., self-propulsion strengths) have recently been discussed to investigate phase behavior unique to multicomponent active matter~\cite{Kolb2020, deCastro2021a, deCastro2021b, RojasVega2023, Lauersdorf2025}.
In the following, as one of the simplest setups, we consider a binary mixture of active particles with distinct activities on a lattice, where any site is occupied by either type of particle (Fig.~\ref{fig_model}).
Motivated by lattice solution models for equilibrium phase separation (e.g., the Flory-Huggins model~\cite{Rubinstein2003}), we regard the first type as solute and the second type as solvent.
In a limit where solvent activity is zero, solutes and solvents are analogous to self-propelled particles and empty space, respectively, in dry active matter.
Furthermore, in another limit where activities of solute and solvent are equal, we can theoretically access a condition where solute-rich and solute-poor regions are symmetric upon exchanging the roles of solute and solvent; this symmetry, as discussed in the following sections, reduces the complexity to gain insight into nonequilibrium phase behavior.
In contrast, such a symmetric condition is difficult to realize in typical active particle systems, where particle-rich and particle-poor regions are asymmetric.

Specifically, we define an active binary mixture model as follows (Fig.~\ref{fig_model}).
Each site of a rectangular lattice with size $(L_x, L_y)$ is occupied by a solute or solvent with polarity $p \in \{ \rightarrow, \uparrow, \leftarrow, \downarrow \}$, which represents the direction of activity (i.e., self-propulsion).
We assume two kinds of stochastic dynamics: (i) onsite rotation of polarity by $+90^\circ$ or $-90^\circ$, each at rate $\gamma$, and (ii) exchange of a solute and a solvent adjacent to each other at a rate depending on the configuration as follows; (ii-1) passive exchange occurs at rate $1$ (i.e., used as the unit of rate) regardless of the polarity or direction of motion; (ii-2) active exchange occurs at rate $\varepsilon_\mathrm{solute}$ ($\varepsilon_\mathrm{solvent}$) if the polarity of solute (solvent) is aligned with its moving direction; in particular, for active exchange with both the solute polarity and solvent polarity aligned with their directions of motion, the corresponding total rate is $\varepsilon_\mathrm{solute} + \varepsilon_\mathrm{solvent}$.
The active exchange represents the self-propelled motion of solute and/or solvent, which violates the detailed balance.
These dynamical rules are summarized in Fig.~\ref{fig_model}.

We use the lattice constant as the unit of length, assume periodic boundary conditions along the $x$ and $y$ axes, and write the mean density (i.e., occupancy) of solutes as $\rho \in (0, 1)$.
Since the passive exchange rate is used as the unit of rate, the dimensionless groups can be taken as $\{ \gamma, \varepsilon_\mathrm{solute}, \varepsilon_\mathrm{solvent}, \rho \}$.
For the numerical simulations in the following, we fix the rotation rate as $\gamma = 0.1$ and examine the change in collective behaviors when we change the solute activity $\varepsilon_\mathrm{solute}$, solvent activity $\varepsilon_\mathrm{solvent}$, and solute density $\rho$ (see Appendix~\ref{app_implementation_of_stochastic_dynamics} for the implementation of stochastic dynamics).

We stress that neither the exchange of two solutes nor the exchange of two solvents is allowed in the present model.
Though such exchange dynamics between the same species has been considered in active mixture models previously~\cite{deCastro2021a, deCastro2021b}, we choose the simplified setup with only the exchange between a solute and a solvent to investigate phase behavior by theoretically connecting the two limits mentioned above and explained more below.

In the limit of no solvent activity ($\varepsilon_\mathrm{solvent} = 0$), this model is equivalent to the active lattice gas~\cite{Thompson2011, Soto2014, Whitelam2018, Kourbane-Houssene2018, Partridge2019}, where solutes with large solute activity $\varepsilon_\mathrm{solute}$ undergo self-aggregation, i.e., active phase separation.
By introducing $\varepsilon_\mathrm{solvent} > 0$, we expect that solvents will also tend to aggregate due to their activity, which might help phase separation cooperatively with solute activity; however, the effect of solvent activity is not so simple, as we discuss below.
In the symmetric limit ($\varepsilon_\mathrm{solute} = \varepsilon_\mathrm{solvent}$), the solutes and solvents behave equivalently (i.e., the dynamical rule does not change if we switch the roles of solute and solvent), and thus the critical density for phase separation should be located at $\rho = 0.5$ from symmetry.

In the following, we mainly focus on the case with $\varepsilon_\mathrm{solute} \geq \varepsilon_\mathrm{solvent}$ and investigate the phase behavior by numerical simulations (Sec.~\ref{sec_bubble_formation}) and mean-field theory (Sec.~\ref{sec_mean-field_theory}).
Then we consider the symmetric limit ($\varepsilon_\mathrm{solute} = \varepsilon_\mathrm{solvent}$) and examine the critical properties for active phase separation (Sec.~\ref{sec_criticality_under_bubble_supression}).

\begin{figure*}[t]
    \centering
    \includegraphics[scale=1]{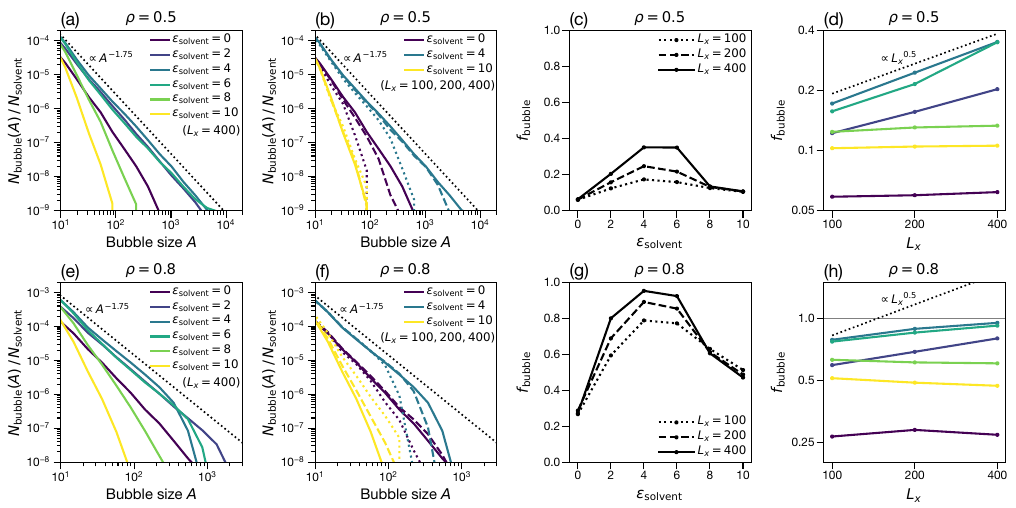}
    \caption{Bubble formation controlled by solvent activity.
    (a, b) Bubble size distribution divided by the total number of solvents, $N_\mathrm{bubble} (A) / N_\mathrm{solvent}$, at $\rho = 0.5$ and $\varepsilon_\mathrm{solute} = 10$.
    The brightness suggests solvent activity $\varepsilon_\mathrm{solvent}$, and the line style indicates the system size [$(L_x, L_y) = (100, 50)$ (dotted lines), $(L_x, L_y) = (200, 100)$ (dashed lines), and $(L_x, L_y) = (400, 200)$ (solid lines)].
    For moderate activity ($\varepsilon_\mathrm{solvent} = 2, 4, 6$) and large enough system sizes, the curves approximately follow a power-law scaling $\sim A^{-1.75}$ (black dotted line).
    (c, d) Bubble fraction $f_\mathrm{bubble}$ at $\rho = 0.5$ as a function of (c) $\varepsilon_\mathrm{solvent}$ or (d) $L_x$, where brightness suggests $\varepsilon_\mathrm{solvent}$ as in (a).
    For moderate $\varepsilon_\mathrm{solvent}$, the bubble fraction follows an approximate scaling: $f_\mathrm{bubble} \sim {L_x}^\mathrm{0.5}$ [black dotted line in (d)].
    (e-h) The corresponding plots at $\rho = 0.8$ and $\varepsilon_\mathrm{solute} = 10$.}
    \label{fig_bubble}
\end{figure*}

\section{Bubble formation controlled by asymmetry of activity}
\label{sec_bubble_formation}

\subsection{Qualitative phase behavior}

In Fig.~\ref{fig_phase_diagram}, we show qualitative steady-state phase diagrams for active phase separation in the $(\varepsilon_\mathrm{solute}, \varepsilon_\mathrm{solvent})$ plane for $(L_x, L_y) = (100, 50)$ at (a) $\rho = 0.5$ and (b) $\rho = 0.8$.
The brightness of each heatmap indicates $\Delta \rho$, the shifted difference in solute density between the dense and dilute regions, which represents the propensity for phase separation (see Appendix~\ref{app_numerical_qualitative_phase_diagram} for details).
Note that nonzero $\Delta \rho$ does not necessarily suggest phase separation.
We also present typical snapshots at several sets of activities $(\varepsilon_\mathrm{solute}, \varepsilon_\mathrm{solvent})$ for $(L_x, L_y) = (100, 50)$ and $(L_x, L_y) = (400, 200)$.

Figure~\ref{fig_phase_diagram} suggests that phase separation is induced by solute activity and/or solvent activity.
Furthermore, as represented by the snapshot at $(\varepsilon_\mathrm{solute}, \varepsilon_\mathrm{solvent}) = (10, 4)$ in Fig.~\ref{fig_phase_diagram}(a), moderate asymmetry in solute and solvent activities enhances formation of solvent bubbles (shown with white color) within a large cluster of solutes (shown with black color); conversely, at $(\varepsilon_\mathrm{solute}, \varepsilon_\mathrm{solvent}) = (4, 10)$, bubbles of solutes form within a large cluster of solvents.
At a higher density ($\rho = 0.8$), the snapshots in Fig.~\ref{fig_phase_diagram}(b) illustrate that bubble formation is enhanced in a similar parameter region.

In Fig.~\ref{fig_phase_diagram}(c), we show the time evolution of a typical configuration in the steady state at $\rho = 0.5$ and $(\varepsilon_\mathrm{solute}, \varepsilon_\mathrm{solvent}) = (10, 4)$.
We find that bubbles persistently form and grow within the bulk solute-rich phase and merge into the bulk solvent-rich phase.
This type of nonstationary bubble dynamics in active phase separation has also been found in active particle models~\cite{Shi2020} and coarse-grained models~\cite{Tjhung2018}.

In the symmetric limit ($\varepsilon_\mathrm{solute} = \varepsilon_\mathrm{solvent}$), which corresponds to the diagonal line in Figs.~\ref{fig_phase_diagram}(a) and \ref{fig_phase_diagram}(b), we find that bubble formation is suppressed.
This is natural considering the result of a coarse-grained model, AMB+~\cite{Tjhung2018, Fausti2024, Cates2024}, in the following way.
The $\zeta$ term in AMB+, which breaks time-reversal symmetry as well as the symmetry between dense and dilute phases, generates a nonlocal chemical potential as a potential origin of persistent bubble formation~\cite{Tjhung2018}.
Since the symmetric limit considered here respects the symmetry between dense (i.e., solute-rich) and dilute (i.e., solvent-rich) phases, the counterpart of the $\zeta$ term and associated bubble formation are expected to be suppressed.

\subsection{Power-law size distribution of enhanced bubbles}

To examine the evolution of solvent bubbles as a function of $\varepsilon_\mathrm{solvent}$ more quantitatively, we consider deeply phase-separated states with large solute activity ($\varepsilon_\mathrm{solute} = 10$) at a fixed density ($\rho = 0.5$).
We plot the bubble size distribution for a fixed system size $(L_x, L_y) = (400, 200)$ in Fig.~\ref{fig_bubble}(a).
The vertical axis is $N_\mathrm{bubble} (A) / N_\mathrm{solvent}$, where $N_\mathrm{bubble} (A)$ is the ensemble-averaged total number of bubbles of size (i.e., area) $A$, and $N_\mathrm{solvent} = (1 - \rho) L_x L_y$ is the total number of solvents.
The bubbles are defined as clusters of solvents, excluding the largest cluster, which is regarded as the bulk solvent-rich phase (see Appendix~\ref{app_bubble_formation} for details).
As solvent activity increases from $\varepsilon_\mathrm{solvent} = 0$ to $\varepsilon_\mathrm{solvent} = 6$, the bubble size distribution becomes broader [Fig.~\ref{fig_bubble}(a)], consistent with the enhanced bubbles observed in the snapshots [Fig.~\ref{fig_phase_diagram}(a)].
Furthermore, for moderate asymmetry of activity ($\varepsilon_\mathrm{solvent} = 2, 4, 6$), we find an approximate power-law distribution:
\begin{equation}
    \frac{N_\mathrm{bubble} (A)}{N_\mathrm{solvent}} \sim A^{-\alpha}
    \label{eq_bubble_size_dist_scaling}
\end{equation}
with $\alpha \approx 1.75$, comparable to the power-law scaling observed in simulations of active lattice gas and active Brownian particles with larger system sizes~\cite{Shi2020}.
As shown in Fig.~\ref{fig_bubble}(b), upon increasing the system size from $(L_x, L_y) = (100, 50)$ (dotted lines) to $(L_x, L_y) = (400, 200)$ (solid lines), the distribution for $\varepsilon_\mathrm{solvent} = 4$ approaches the power-law scaling regime more quickly than that for $\varepsilon_\mathrm{solvent} = 0$ (i.e., the case equivalent to active lattice gas).

As discussed previously~\cite{Shi2020, Nakano2024}, the power-law distribution~\eqref{eq_bubble_size_dist_scaling} with $\alpha \approx 1.75 < 2$ means that the bubble fraction,
\begin{equation}
    f_\mathrm{bubble} := \frac{1}{N_\mathrm{solvent}} \sum_A A N_\mathrm{bubble} (A),
    \label{eq_bubble_fraction}
\end{equation}
should grow algebraically with $L_x$:
\begin{equation}
    f_\mathrm{bubble} \sim {L_x}^{2(2 - \alpha)},
    \label{eq_bubble_fraction_scaling}
\end{equation}
where we assume that the aspect ratio $L_x / L_y$ is fixed and the upper limit of the summation in Eq.~\eqref{eq_bubble_fraction} is proportional to $L_x L_y$~\cite{Nakano2024}.
Note that, since the lattice constant is used as the unit of length, $A$ is equal to the total particle number contained in a bubble, leading to $0 \leq f_\mathrm{bubble} < 1$.
As demonstrated in Fig.~\ref{fig_bubble}(c), $f_\mathrm{bubble}$ is indeed enhanced for $\varepsilon_\mathrm{solvent} = 2, 4, 6$ as the system size increases.
Moreover, as shown with the log-log plot in Fig.~\ref{fig_bubble}(d), $f_\mathrm{bubble}$ approximately follows the predicted scaling~\eqref{eq_bubble_fraction_scaling} with $2 (2 - \alpha) \approx 0.5$, consistent with $\alpha \approx 1.75$ in the distribution~\eqref{eq_bubble_size_dist_scaling}.

At larger solvent activities ($\varepsilon_\mathrm{solvent} = 8, 10$), which are comparable to the solute activity ($\varepsilon_\mathrm{solute} = 10$), the bubble size distribution is narrower and does not follow the power-law scaling, as shown in Fig.~\ref{fig_bubble}(a).
This presumably reflects the suppression of bubbles by symmetry between solutes and solvents.
Consistently, the bubble fraction $f_\mathrm{bubble}$ is almost independent of the system size [Figs.~\ref{fig_bubble}(c) and \ref{fig_bubble}(d)], in contrast to the cases with moderate asymmetry ($\varepsilon_\mathrm{solvent} = 2, 4, 6$).

To examine the generality of the observed effects of solvent activity, we also consider systems with a higher solute density ($\rho = 0.8$) [see Fig.~\ref{fig_phase_diagram}(b) for snapshots].
As shown in Figs.~\ref{fig_bubble}(e) and \ref{fig_bubble}(f), the bubble size distribution follows the approximate power-law scaling for $\varepsilon_\mathrm{solvent} = 2, 4, 6$ in a similar way as observed for $\rho = 0.5$.
Further, bubble formation is enhanced for $\varepsilon_\mathrm{solvent} = 2, 4, 6$ as the system size increases, as suggested from Figs.~\ref{fig_bubble}(f) and \ref{fig_bubble}(g).
On the other hand, Fig.~\ref{fig_bubble}(h) shows that the scaling law for $f_\mathrm{bubble}$ [Eq.~\eqref{eq_bubble_fraction_scaling}] is violated, and $f_\mathrm{bubble}$ is almost saturated near $1$ especially at $\varepsilon_\mathrm{solvent} = 4$.
At larger solvent activities ($\varepsilon_\mathrm{solvent} = 8, 10$), bubbles are suppressed [Figs.~\ref{fig_bubble}(e)-\ref{fig_bubble}(h)] in a similar way as observed for $\rho = 0.5$.

Since $0 \leq f_\mathrm{bubble} < 1$ by definition, $f_\mathrm{bubble}$ is expected to converge to a value less than or equal to $1$ as $L_x L_y \to \infty$.
When the power-law distribution appears in finite systems, the asymptotic phase behavior for $L_x L_y \to \infty$ should be classified into either bubbly phase separation~\cite{Tjhung2018, Fausti2024} or microphase separation~\cite{Tjhung2018, Shi2020}.
In the case of bubbly phase separation, the largest solvent cluster scales with the system size, leading to $f_\mathrm{bubble} \to \mathrm{const.} < 1$ for $L_x L_y \to \infty$.
In the case of microphase separation, all the solvent clusters exist as bubbles with finite sizes that do not scale with the system size, leading to $f_\mathrm{bubble} \to 1$ for $L_x L_y \to \infty$.
We need larger-scale simulations to confirm the $\rho$ and $\varepsilon_\mathrm{solvent}$ dependence of the asymptotic phase behavior.

Overall, in the deeply phase-separated states caused by large solute activity, we can enhance or suppress bubble formation by moderate or large solvent activity, respectively.
From symmetry, the discussions above are also applied to systems with $\varepsilon_\mathrm{solute} \leq \varepsilon_\mathrm{solvent}$ by exchanging the roles of solutes and solvents.

\begin{figure}[t]
    \centering
    \includegraphics[scale=1]{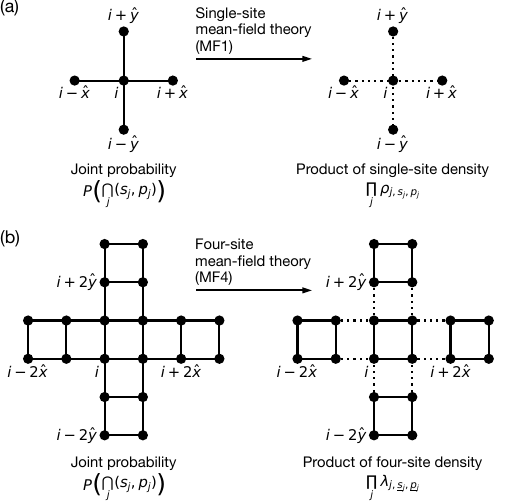}
    \caption{Schematic explanation of (a) single-site mean-field theory and (b) four-site mean-field theory.
    Black dots indicate lattice sites, and dotted lines represent the neglected correlations between (a) adjacent sites or (b) adjacent clusters.}
    \label{fig_mean-field_schematic}
\end{figure}

\begin{figure*}[t]
    \centering
    \includegraphics[scale=1]{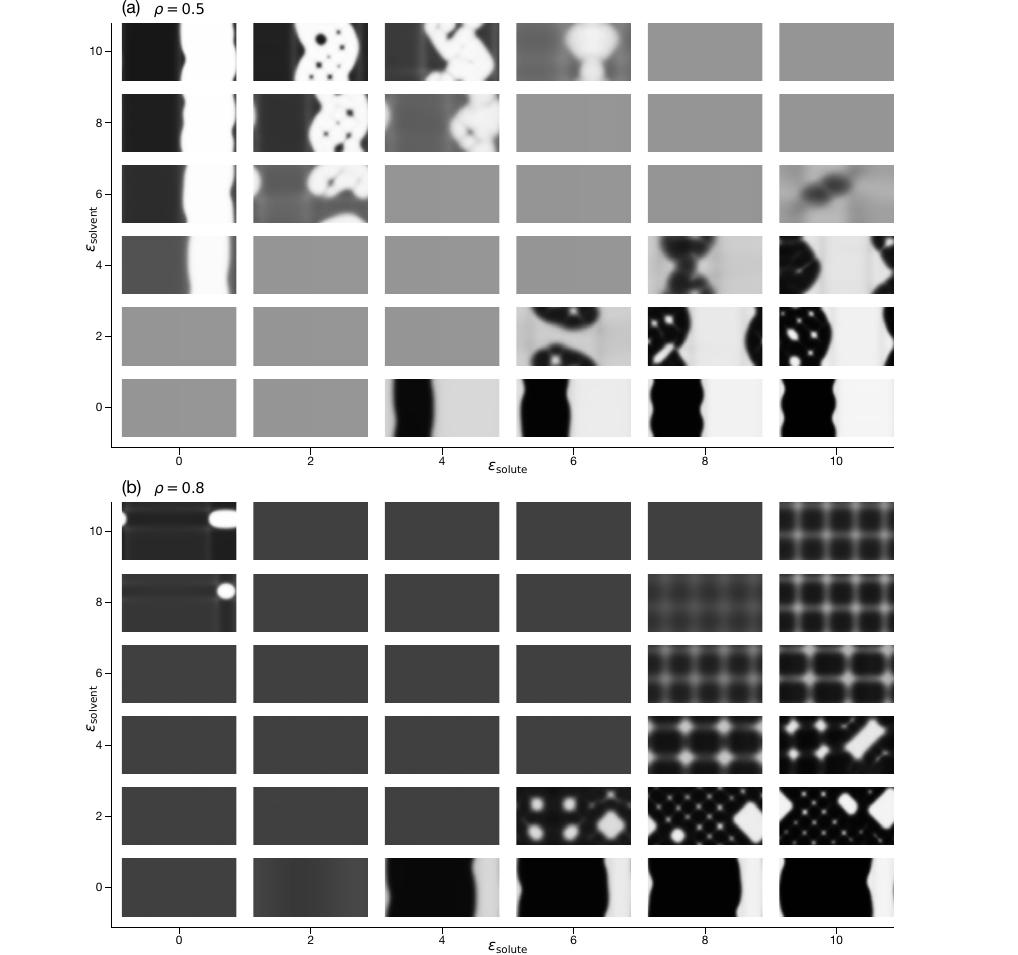}
    \caption{Phase behavior in mean-field dynamics.
    (a, b) Typical snapshots for different sets of $(\varepsilon_\mathrm{solute}, \varepsilon_\mathrm{solvent})$ with $(L_x, L_y) = (100, 50)$ at (a) $\rho = 0.5$ or (b) $\rho = 0.8$.
    Grayscale represents solute density $\rho_{i, \mathrm{solute}}$ (white for $\rho_{i, \mathrm{solute}} = 0$ and black for $\rho_{i, \mathrm{solute}} = 1$).}
    \label{fig_mean-field_snapshots}
\end{figure*}

\begin{figure*}[t]
    \centering
    \includegraphics[scale=1]{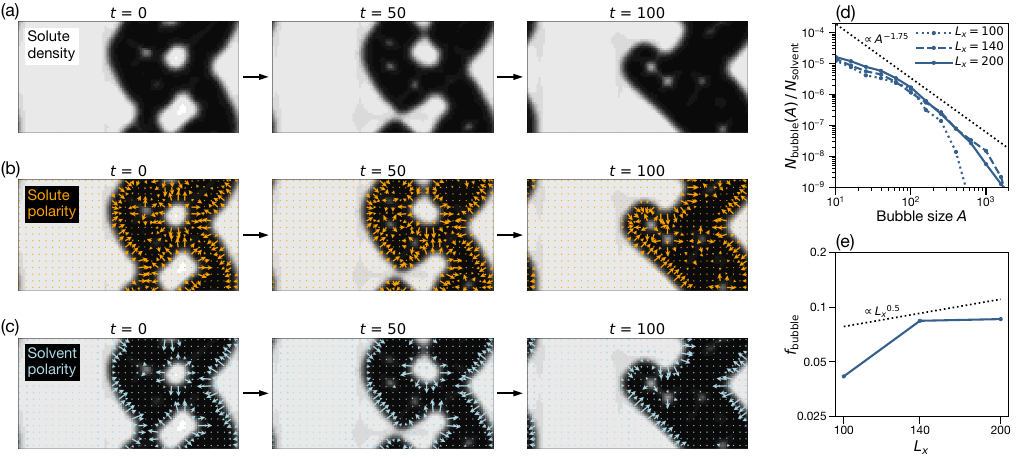}
    \caption{Nonstationary bubble formation in mean-field dynamics.
    (a-c) Typical time evolution of bubbles.
    Grayscale indicates solute density $\rho_{i, \mathrm{solute}}$.
    Arrows in (b) orange and (c) light blue represent solute polarity density $\bm{w}_{i, \mathrm{solute}}$ and solvent polarity density $\bm{w}_{i, \mathrm{solvent}}$, respectively, shown in arbitrary units.
    (d) Bubble size distribution divided by the total number of solvents, $N_\mathrm{bubble} (A) / N_\mathrm{solvent}$, for three system sizes [$(L_x, L_y) = (100, 50), (140, 70), (200, 100)$], shown with a power-law function $\propto A^{-1.75}$ (black dotted line).
    (e) Bubble fraction $f_\mathrm{bubble}$, shown with a power-law function $\propto {L_x}^{0.5}$ (black dotted line).
    For all panels, we set $\rho = 0.5$ and $(\varepsilon_\mathrm{solute}, \varepsilon_\mathrm{solvent}) = (10, 3)$.}
    \label{fig_mean-field_bubble}
\end{figure*}

\section{Mean-field theory}
\label{sec_mean-field_theory}

To capture the observed phase behavior within reduced models, we employ mean-field theory~\cite{Solon2013, Solon2015aim, Curatolo2016, Zakine2024}.
A motivation for applying mean-field theory is to gain an approximate yet analytical insight into the observed phenomena.
In the following, we provide a useful starting point toward a systematic understanding of phase separation phenomena in the active binary mixture model, especially bubble formation, by mean-field approximations combined with linear stability analysis.

\subsection{Single-site mean-field theory}
\label{sec_single-site_mean-field_theory}

As a simple way for model reduction, we consider single-site mean-field theory.
We define $\rho_{i, s, p} (t)$ as the marginal probability that site $i$ is occupied by species $s$ ($\in \{ \mathrm{solute}, \mathrm{solvent} \}$) with polarity $p$ at time $t$.
We can also regard $\rho_{i, s, p} (t)$ as the ensemble-averaged local density of species $s$ with polarity $p$ at site $i$ and time $t$.
As shown in Fig.~\ref{fig_mean-field_schematic}(a), by neglecting microscopic correlations across sites in a similar way as applied previously~\cite{Solon2013, Solon2015aim, Curatolo2016, Zakine2024}, we obtain a dynamical mean-field equation for $\rho_{i, s, p} (t)$:
\begin{equation}
    \partial_t \rho_{i, s, p} = f_{i, s, p}^{(p)} + f_{i, s, p}^{(a)} + f_{i, s, p}^{(r)}.
    \label{eq_mean_field_1site}
\end{equation}

The first term $f^{(p)}_{i, s, p}$ comes from the passive exchange of a solute and a solvent:
\begin{equation}
    f^{(p)}_{i, s, p} = \sum_{l \in \{ \pm \hat{x}, \pm \hat{y} \}} \sum_{p'} (\rho_{i, \bar{s}, p'} \rho_{i + l, s, p} - \rho_{i + l, \bar{s}, p'} \rho_{i, s, p}),
\end{equation}
where $\bar{s}$ suggests the species opposite to $s$ (e.g., $\overline{\mathrm{solute}} = \mathrm{solvent}$), $\hat{x}$ and $\hat{y}$ are the unit translations in the $x$ and $y$ directions, respectively, and the polarity indices $(\rightarrow, \uparrow, \leftarrow, \downarrow)$ are identified as $(+\hat{x}, +\hat{y}, -\hat{x}, -\hat{y})$.
The second term $f^{(a)}_{i, s, p}$ derives from the active exchange of a solute and a solvent:
\begin{align}
    f^{(a)}_{i, s, p} &= \sum_{p'} \varepsilon_s (\rho_{i, \bar{s}, p'} \rho_{i - p, s, p} - \rho_{i + p, \bar{s}, p'} \rho_{i, s, p}) \nonumber \\
    &\quad + \sum_{p'} \varepsilon_{\bar{s}} (\rho_{i, \bar{s}, p'} \rho_{i + p', s, p} - \rho_{i - p', \bar{s}, p'} \rho_{i, s, p}).
\end{align}
The last term $f^{(r)}_{i, s, p}$ corresponds to the rotation of polarity:
\begin{equation}
    f^{(r)}_{i, s, p} = \gamma (\rho_{i, s, R p} + \rho_{i, s, R^{-1} p} - 2 \rho_{i, s, p}),
\end{equation}
where $R$ and $R^{-1}$ represent counterclockwise and clockwise rotations by $90^\circ$, respectively [e.g., $R (+\hat{y}) = -\hat{x}$].

\subsection{Phase behavior in mean-field dynamics}
\label{sec_phase_behavior_in_mean-field_dynamics}

In Fig.~\ref{fig_mean-field_snapshots}(a), we present typical snapshots of the solute density, $\rho_{i, \mathrm{solute}} := \sum_p \rho_{i, \mathrm{solute}, p}$, for $(L_x, L_y) = (100, 50)$ in the $(\varepsilon_\mathrm{solute}, \varepsilon_\mathrm{solvent})$ plane at $\rho = 0.5$.
Each snapshot is obtained by numerical integration of Eq.~\eqref{eq_mean_field_1site} until $t = 2000$ using the Euler method with time step $dt = 0.02$, where the initial state is a fully phase-separated configuration with random polarity [see Fig.~\ref{appfig_initial_config}(b)].
We find that the mean-field dynamics qualitatively reproduces the enhancement of bubble formation by moderate asymmetry of activity as seen in the original model [Fig.~\ref{fig_phase_diagram}(a)].

As demonstrated in Fig.~\ref{fig_mean-field_bubble}(a) for $(\varepsilon_\mathrm{solute}, \varepsilon_\mathrm{solvent}) = (10, 3)$, we find that nonstationary bubble dynamics appears for a certain range of activity parameters, even though the mean-field equation [Eq.~\eqref{eq_mean_field_1site}] is deterministic.
Bubbles persistently form and grow within the bulk solute-rich phase and merge into the bulk solvent-rich phase, in a similar way as observed in stochastic simulations of the original model [Fig.~\ref{fig_phase_diagram}(c)].
We plot the corresponding bubble size distribution in Fig.~\ref{fig_mean-field_bubble}(d) (see Appendix~\ref{app_bubble_formation_in_mean-field_dynamics} for details), which suggests that larger bubbles tend to appear in larger systems, as seen in the original model [Fig.~\ref{fig_bubble}(b)].
However, the distribution saturates at relatively small system sizes, such as $(L_x, L_y) = (200, 100)$.
Also, the power laws in the distribution [i.e., $N_\mathrm{bubble} (A) / N_\mathrm{solvent} \sim A^{-1.75}$] and bubble fraction (i.e., $f_\mathrm{bubble} \sim {L_x}^{0.5}$) are not so clear [Figs.~\ref{fig_mean-field_bubble}(d) and \ref{fig_mean-field_bubble}(e)], compared to the obaservations in the original model [Figs.~\ref{fig_bubble}(b) and \ref{fig_bubble}(d)].
Such discrepancies may come from the absence of stochasticity in the mean-field dynamics, since stochasticity should increase the nucleation rate of bubbles and affect the bubble size distribution.

To consider the mechanism of enhanced bubble formation within mean-field theory, we examine the polarity density of solutes and solvents, defined as $\bm{w}_{i, s} := \sum_p p \rho_{i, s, p}$, for $(\varepsilon_\mathrm{solute}, \varepsilon_\mathrm{solvent}) = (10, 3)$.
As illustrated by the colored arrows in Figs.~\ref{fig_mean-field_bubble}(b) and \ref{fig_mean-field_bubble}(c), the polarity densities of solutes and solvents are aligned near the interface between the solute-rich and solvent-rich phases.
First, the polarity of solutes is oriented toward the solute-rich phase, in a similar way as observed in phase separation of active particles~\cite{Hermann2019, Hermann2020, Omar2023}.
This alignment of solute polarity derives from solute activity and exclusion, and is interpreted to stabilize the large cluster of solutes against diffusion.
Second, the polarity density of solvents is also oriented toward the solute-rich phase, which reflects asymmetry in activity and is presumably due to the particle exchange driven by solvent activity near the interface.
This alignment of solvent polarity is interpreted to expand bubbles within the solute-rich phase, leading to the enhancement of bubble formation.

In Fig.~\ref{fig_mean-field_snapshots}(b), we show typical snapshots in the case of $\rho = 0.8$.
We find that bubble formation is enhanced for moderate asymmetry of activity, such as $(\varepsilon_\mathrm{solute}, \varepsilon_\mathrm{solvent}) = (10, 2)$, in a similar way as observed in the original model [Fig.~\ref{fig_phase_diagram}(b)].
The periodic patterns seen in the mean-field dynamics [Fig.~\ref{fig_mean-field_snapshots}(b)] suggest the propensity for microphase separation at high density, though bubbles exhibit more fluctuating behaviors in the original stochastic model [Fig.~\ref{fig_phase_diagram}(b)].

The mean-field dynamics does not reproduce macroscopic phase separation near the symmetric limit ($\varepsilon_\mathrm{solute} = \varepsilon_\mathrm{solvent}$), which is observed in the original model [Figs.~\ref{fig_phase_diagram}(a) and \ref{fig_phase_diagram}(b)]; only the homogeneous state appears for $\rho = 0.5$ [Fig.~\ref{fig_mean-field_snapshots}(a)], and the homogeneous state or periodic pattern appears for $\rho = 0.8$ [Fig.~\ref{fig_mean-field_snapshots}(b)].
This indicates a qualitative shortcoming in single-site mean-field theory.
In particular, for $\rho = 0.5$, a critical point for macroscopic phase separation should exist on the line $\varepsilon_\mathrm{solute} = \varepsilon_\mathrm{solvent}$, highlighting the need to refine mean-field theory, which we conduct in Sec.~\ref{sec_four-site_mean-field_theory}.

\begin{figure}[t]
    \centering
    \includegraphics[scale=1]{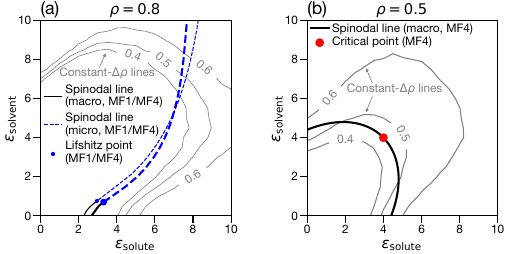}
    \caption{Mean-field spinodal line and critical point.
    (a) Spinodal line for macroscopic phase separation (black solid line) or microphase separation (blue dashed line) and the Lifshitz point (blue dot) at $\rho = 0.8$.
    The thin and thick lines are obtained by single-site and four-site mean-field theories, respectively.
    (b) Spinodal line (black solid line) and critical point (red dot) for macroscopic phase separation at $\rho = 0.5$, obtained by four-site mean-field theory.
    In each panel, the constant-$\Delta \rho$ lines (thin gray lines) are reproduced from Fig.~\ref{fig_phase_diagram} for reference.}
    \label{fig_mean-field_spinodal}
\end{figure}

\subsection{Linear stability analysis}
\label{sec_linear_stability_analysis}

Motivated by the mean-field dynamics reproducing the enhancement of bubbles by moderate activity asymmetry, we apply linear stability analysis to Eq.~\eqref{eq_mean_field_1site} around the homogeneous state [i.e., $\rho_{i, \mathrm{solute}, p} = \rho / 4$ and $\rho_{i, \mathrm{solvent}, p} = (1 - \rho) / 4$ for any $i$ or $p$].
From Eq.~\eqref{eq_mean_field_1site}, we derive the linearized dynamical equation for the Fourier mode of fluctuations, $\bm{\rho}_{\bm{k}} (t)$:
\begin{equation}
    \partial_t \bm{\rho}_{\bm{k}} = R_{\bm{k}}^\mathrm{MF1} \bm{\rho}_{\bm{k}}.
    \label{eq_mean_field_1site_linearized_ratemat}
\end{equation}
Here, $\bm{\rho}_{\bm{k}}$ is an $8$-dimensional vector ($[\bm{\rho}_{\bm{k}}]_{s, p} := \rho_{\bm{k}, s, p} := \sum_j e^{-i \bm{k} \cdot j} \rho_{j, s, p}$), and $R_{\bm{k}}^\mathrm{MF1}$ is an $8 \times 8$ matrix (see Appendix~\ref{app_linear_stability_analysis_for_single-site_mean-field_theory} for details).

Since $\sum_{s, p} \rho_{\bm{k}, s, p} = 0$ holds for each $\bm{k}$, which is derived from $\sum_{s, p} \rho_{i, s, p} = 1$ for any $i$, the total number of independent variables in Eq.~\eqref{eq_mean_field_1site_linearized_ratemat} is $7$ for each $\bm{k}$.
Thus, $R_{\bm{k}}^\mathrm{MF1}$ has one trivial zero eigenvalue for each $\bm{k}$.
After excluding this zero eigenvalue, we define $r_{\bm{k}}^\mathrm{MF1}$ as the eigenvalue of $R_{\bm{k}}^\mathrm{MF1}$ with the largest real part.
If $\mathrm{Re} \, r_{\bm{k}}^\mathrm{MF1} > 0$ for some $\bm{k}$, the homogeneous state is unstable against fluctuation with wavevector $\bm{k}$.
On the other hand, if $\mathrm{Re} \, r_{\bm{k}}^\mathrm{MF1} < 0$ for any $\bm{k} \neq \bm{0}$, the homogeneous state is linearly stable.
Note that $r_{\bm{0}}^\mathrm{MF1} = 0$ always holds due to the particle number conservation (i.e., $\sum_{i, p} \rho_{i, \mathrm{solute}, p} = \rho L^2$).

In Fig.~\ref{fig_mean-field_spinodal}(a) with thin black solid and thin blue dashed lines, we show the spinodal line (i.e., stability limit of the homogeneous state) in the $(\varepsilon_\mathrm{solute}, \varepsilon_\mathrm{solvent})$ plane at $\rho = 0.8$.
To obtain the spinodal line, at which $\max_{\bm{k}} \mathrm{Re} \, r_{\bm{k}}^\mathrm{MF1}$ shifts from zero to a positive value, we numerically maximized $\mathrm{Re} \, r_{\bm{k}}^\mathrm{MF1}$ with a Python package (scipy.optimize.differential\_evolution~\cite{Virtanen2020}) using the default parameters.
We find two regions of the spinodal line: one toward macroscopic phase separation (i.e., instability with $\bm{k} \to \bm{0}$, shown with a thin black solid line), and the other toward microphase separation (i.e., instability with $\bm{k} \neq \bm{0}$, shown with a thin blue dashed line).
These two regions merge at the Lifshitz point~\cite{Chaikin1995}, indicated by a small blue dot.
For comparison, we also show the constant-$\Delta \rho$ lines reproduced from Fig.~\ref{fig_phase_diagram}(b) with thin gray lines.
The results suggest that the tendency toward microphase separation induced by solvent activity is explained by linear instability, and the constant-$\Delta \rho$ lines (i.e., transition lines) are qualitatively captured by the spinodal line.
Note that the spinodal lines do not follow the constant-$\Delta \rho$ lines on the upper side in the $(\varepsilon_\mathrm{solute}, \varepsilon_\mathrm{solvent})$ plane, where the transition can happen only through nucleation, which cannot be explained by linear stability analysis.

To gain further insight into microphase separation caused by solvent activity, we focus on the functional dependence of $r_{\bm{k}}^\mathrm{MF1}$ on $\varepsilon_\mathrm{solute}$, $\varepsilon_\mathrm{solvent}$, and $\rho$.
With the adiabatic approximation for fast variables~\cite{Speck2014, Speck2015} and an assumption that the diagonal modes (i.e., $k_x = \pm k_y$) are most unstable (see Appendix~\ref{app_linear_stability_analysis_for_single-site_mean-field_theory} for details), we obtain the analytical expression of $r_{\bm{k}}^\mathrm{MF1}$ for small wavenumber $|\bm{k}|$:
\begin{equation}
    r_{\bm{k}}^\mathrm{MF1} = c_2^\mathrm{MF1} |\bm{k}|^2 + c_{4+}^\mathrm{MF1} |\bm{k}|^4 + O(|\bm{k}|^6),
\end{equation}
where
\begin{widetext}
\begin{align}
    c_2^\mathrm{MF1} & = -\left( 1 + \frac{\varepsilon_\mathrm{solute} + \varepsilon_\mathrm{solvent}}{4} \right) + \frac{{\varepsilon_\mathrm{solute}}^2 (1 - \rho) (2 \rho - 1)}{4 \gamma} - \frac{{\varepsilon_\mathrm{solvent}}^2 \rho (2 \rho - 1)}{4 \gamma}, \label{eq_c2_mf1} \\
    c_{4+}^\mathrm{MF1} & = \frac{1}{24} \left( 1 + \frac{\varepsilon_\mathrm{solute} + \varepsilon_\mathrm{solvent}}{4} \right) - \frac{{\varepsilon_\mathrm{solute}}^2 (1 - \rho) (2 \rho - 1)}{24 \gamma} + \frac{{\varepsilon_\mathrm{solvent}}^2 \rho (2 \rho - 1)}{24 \gamma} \nonumber \\
    & \quad - \frac{{\varepsilon_\mathrm{solute}}^2}{8 \gamma^2} \left( 1 + \frac{\varepsilon_\mathrm{solute} + \varepsilon_\mathrm{solvent}}{4} \right) (1 - \rho)^2 (2 \rho - 1) + \frac{{\varepsilon_\mathrm{solvent}}^2}{8 \gamma^2} \left( 1 + \frac{\varepsilon_\mathrm{solute} + \varepsilon_\mathrm{solvent}}{4} \right) \rho^2 (2 \rho - 1). \label{eq_c4p_mf1}
\end{align}
\end{widetext}
The spinodal line for macroscopic phase separation [thin black solid line in Fig.~\ref{fig_mean-field_spinodal}(a)] corresponds to the condition for $c_2^\mathrm{MF1} = 0$, and the Lifshitz point [small blue dot in Fig.~\ref{fig_mean-field_spinodal}(a)] is determined by $c_2^\mathrm{MF1} = 0$ and $c_{4+}^\mathrm{MF1} = 0$.

The expression of $c_2^\mathrm{MF1}$ [Eq.~\eqref{eq_c2_mf1}] suggests distinct effects of solute and solvent activities.
The first term indicates that both solute and solvent activities linearly increase the effective diffusion constant, stabilizing the homogeneous state.
In contrast, the second and third terms highlight the opposite effects of solute and solvent activities at the quadratic order; for $0.5 < \rho < 1$, the solute activity $\varepsilon_\mathrm{solute}$ increases $c_2^\mathrm{MF1}$ and tends to induce instability toward macroscopic phase separation, while the solvent activity $\varepsilon_\mathrm{solvent}$ decreases $c_2^\mathrm{MF1}$ and stabilizes the homogeneous state; for $0 < \rho < 0.5$, the roles of solute and solvent activities are reversed.
Such competition of solute and solvent activities causes the positive slope of the spinodal line in Fig.~\ref{fig_mean-field_spinodal}(a).

The expression of $c_{4+}^\mathrm{MF1}$ [Eq.~\eqref{eq_c4p_mf1}] illuminates further interplay of solute and solvent activities.
For $0.5 < \rho < 1$, the nonlinear contribution of $\varepsilon_\mathrm{solvent}$ shifts $c_\mathrm{4+}^\mathrm{MF1}$ toward positive values as $\varepsilon_\mathrm{solvent}$ increases.
This demonstrates that solvent activity can induce microphase separation (i.e., instability with $\bm{k} \neq \bm{0}$), given that solvent activity concurrently decreases $c_\mathrm{2}^\mathrm{MF1}$ and suppresses macroscopic phase separation as explained above.
For $0 < \rho < 0.5$, the roles of solute and solvent activities are reversed.

Although linear stability analysis qualitatively accounts for the phase behavior at $\rho = 0.8$, it fails to predict a spinodal line at $\rho = 0.5$.
Specifically, Eq.~\eqref{eq_c2_mf1} leads to $c_2^\mathrm{MF1} < 0$ for $\rho = 0.5$ regardless of $(\varepsilon_\mathrm{solute}, \varepsilon_\mathrm{solvent})$, and we numerically confirmed that $\mathrm{Re} \, r_{\bm{k}}^\mathrm{MF1} < 0$ for any $\bm{k} \neq \bm{0}$, indicating that the homogeneous state would remain linearly stable.
This contradicts the presumed existence of the critical point for macroscopic phase separation and an accompanying spinodal line, which motivates us to improve mean-field theory (Sec.~\ref{sec_four-site_mean-field_theory}), as already mentioned at the end of Sec.~\ref{sec_phase_behavior_in_mean-field_dynamics}.

\subsection{Four-site mean-field theory}
\label{sec_four-site_mean-field_theory}

To find the presumed critical point and an accompanying spindal line for $\rho = 0.5$, we refine the mean-field theory.
Instead of the approximation used in Sec.~\ref{sec_single-site_mean-field_theory}, where all the microscopic correlations between neighboring sites are neglected, we incorporate local correlations at the four-site level by considering the four-site density $\lambda_{i, \underline{s}, \underline{p}} (t)$.
Here, $\underline{s} := (s_1, s_2, s_3, s_4)$ and $\underline{p} := (p_1, p_2, p_3, p_4)$ represent the configuration for a square cluster that consists of $i_1 := i$, $i_2 := i + \hat{x}$, $i_3 := i + \hat{x} + \hat{y}$, and $i_4 := i + \hat{y}$, which we call cluster $i$.
More specifically, $\lambda_{i, \underline{s}, \underline{p}} (t)$ is the marginal joint probability that site $i_n$ is occupied by species $s_n$ with polarity $p_n$ for $n \in \{ 1, 2, 3, 4 \}$ at time $t$.
In this subsection, we assume that $L_x$ and $L_y$ are even numbers, and the $x$ and $y$ coordinates of site $i$ are odd numbers.

We derive the dynamical mean-field equation for $\lambda_{i, \underline{s}, \underline{p}} (t)$ by neglecting the correlations across square clusters [Fig.~\ref{fig_mean-field_schematic}(b)]:
\begin{equation}
    \partial_t \lambda_{i, \underline{s}, \underline{p}} = g_{i, \underline{s}, \underline{p}}^{(p, w)} + g_{i, \underline{s}, \underline{p}}^{(a, w)} + g_{i, \underline{s}, \underline{p}}^{(p, n)} + g_{i, \underline{s}, \underline{p}}^{(a, n)} + g_{i, \underline{s}, \underline{p}}^{(r)}.
    \label{eq_mean_field_4site}
\end{equation}
Here, $g_{i, \underline{s}, \underline{p}}^{(p, w)}$ is the contribution from the passive exchange within cluster $i$ (e.g., exchange of solute with polarity $+\hat{y}$ at site $i$ and solvent with polarity $-\hat{y}$ at site $i + \hat{x}$).
Similarly, $g_{i, \underline{s}, \underline{p}}^{(a, w)}$ is the contribution from the active exchange within cluster $i$ (e.g., exchange of solute with polarity $+\hat{x}$ at site $i$ and solvent with polarity $-\hat{y}$ at site $i + \hat{x}$).
Next, $g_{i, \underline{s}, \underline{p}}^{(p, n)}$ and $g_{i, \underline{s}, \underline{p}}^{(a, n)}$ correspond to the passive and active exchanges, respectively, across cluster $i$ and neighboring clusters (e.g., active exchange of solute with polarity $-\hat{x}$ at site $i$ and solvent with polarity $+\hat{y}$ at site $i - \hat{x}$).
Lastly, $g_{i, \underline{s}, \underline{p}}^{(r)}$ represents the rotation of polarity within cluster $i$.
The explicit form of Eq.~\eqref{eq_mean_field_4site} is summarized in Appendix~\ref{app_explicit_form_of_dynamical_mean-field_equation}.

In a similar way as used in Sec.~\ref{sec_linear_stability_analysis}, we first linearize Eq.~\eqref{eq_mean_field_4site} in terms of the fluctuations of $\lambda_{i, \underline{s}, \underline{p}}$ around the homogeneous state, which is determined by solving Eq.~\eqref{eq_mean_field_4site} numerically under the assumption that $\lambda_{i, \underline{s}, \underline{p}}$ is spatially uniform (i.e., independent of $i$).
Then, using $\lambda_{\bm{k}, \underline{s}, \underline{p}}$, the Fourier component of fluctuations with wavevector $\bm{k}$, we obtain the linearized equation in the following form:
\begin{equation}
    \partial_t \bm{\lambda}_{\bm{k}} = R_{\bm{k}}^\mathrm{MF4} \bm{\lambda}_{\bm{k}}.
    \label{eq_mean_field_4site_linearized_ratemat}
\end{equation}
Here, $\bm{\lambda}_{\bm{k}}$ is a $4096$-dimensional vector ($[\bm{\lambda}_{\bm{k}}]_{\underline{s}, \underline{p}} := \lambda_{\bm{k}, \underline{s}, \underline{p}}$), and $R_{\bm{k}}^\mathrm{MF4}$ is a $4096 \times 4096$ matrix, considering that the total number of combinations of $(\underline{s}, \underline{p}) = (s_1, s_2, s_3, s_4, p_1, p_2, p_3, p_4)$ is $2^4 \times 4^4 = 4096$.

Since $\sum_{\underline{s}, \underline{p}} \lambda_{\bm{k}, \underline{s}, \underline{p}} = 0$ holds for each $\bm{k}$, which is derived from $\sum_{\underline{s}, \underline{p}} \lambda_{i, \underline{s}, \underline{p}} = 1$ for any $i$, the total number of independent variables in Eq.~\eqref{eq_mean_field_4site_linearized_ratemat} is $4095$ for each $\bm{k}$.
Thus, $R_{\bm{k}}^\mathrm{MF4}$ has one trivial zero eigenvalue for each $\bm{k}$.
After excluding this zero eigenvalue, we define $r_{\bm{k}}^\mathrm{MF4}$ as the eigenvalue of $R_{\bm{k}}^\mathrm{MF4}$ with the largest real part.
In the same way as explained for $r_{\bm{k}}^\mathrm{MF4}$, the homogeneous state is unstable if $\mathrm{Re} \, r_{\bm{k}}^\mathrm{MF4} > 0$ for some $\bm{k}$, while the homogeneous state is linearly stable if $\mathrm{Re} \, r_{\bm{k}}^\mathrm{MF4} < 0$ for any $\bm{k} \neq \bm{0}$.

We estimate the spinodal line for macroscopic phase separation and microphase separation at $\rho = 0.8$, as plotted with thick black solid and thick blue dashed lines, respectively, in Fig.~\ref{fig_mean-field_spinodal}(a).
Here, to obtain the spinodal line for microphase separation with reasonable computational cost, we maximized $\mathrm{Re} \, r_{\bm{k}}^\mathrm{MF4}$ with a Python package (scipy.optimize.brute~\cite{Virtanen2020}) by assuming that the optimal $\bm{k}$ exists either on the axes (i.e., $k_x = 0$ or $k_y = 0$) or on the diagonals (i.e., $k_x = k_y$ or $k_x = -k_y$).
This assumption should hold at least near the Lifshitz point, and we also numerically verified that it holds far from the Lifshitz point within single-site mean-field theory.
We find that the spinodal lines obtained by single-site and four-site mean-field theories are qualitatively similar [see the thin and thick lines in Fig.~\ref{fig_mean-field_spinodal}(a)].

Examining $r_{\bm{k}}^\mathrm{MF4}$ for $\rho = 0.5$ numerically, we obtain the spinodal line for macroscopic phase separation, as shown with a black line in Fig.~\ref{fig_mean-field_spinodal}(b).
We also identify the critical point on the diagonal line (i.e, $\varepsilon_\mathrm{solute} = \varepsilon_\mathrm{solvent}$), which is marked by a red dot in Fig.~\ref{fig_mean-field_spinodal}(b).
First, the spinodal line based on four-site mean-field theory has a positive slope near $\varepsilon_\mathrm{solvent} = 0$.
This suggests the competition between solute and solvent activities, as discussed more explicitly for $\rho \neq 0.5$ within single-site mean-field theory (Sec.~\ref{sec_linear_stability_analysis}).
Second, the spinodal line reproduces the qualitative feature of the constant-$\Delta \rho$ lines [thin gray lines in Fig.~\ref{fig_mean-field_spinodal}(b)], in contrast to single-site mean-field theory, which fails to predict linear instability at $\rho = 0.5$.
This underlines that local correlations, which are incorporated in four-site mean-field theory but are discarded in single-site mean-field theory, should be crucial in predicting the linear instability toward phase separation and the existence of the critical point in the present case.

At $\rho = 0.5$ in the symmetric limit ($\varepsilon_\mathrm{solute} = \varepsilon_\mathrm{solvent} = \varepsilon$), the critical point activity is obtained numerically as $\varepsilon_c^\mathrm{MF4} \simeq 4.00$, and phase separation is predicted to appear when $\varepsilon > \varepsilon_c^\mathrm{MF4}$.
We expect that $\varepsilon_c^\mathrm{MF4}$ quantitatively deviates from the exact critical point because the latter accounts for correlations spanning all length scales rather than being limited to the four-site correlations.
More importantly, the correlations at large length scales can shift the critical exponents from their mean-field values to those in a specific universality class~\cite{Hohenberg1977, Chaikin1995}, which we discuss in Sec.~\ref{sec_criticality_under_bubble_supression}.

\begin{table}[b]
    \centering
    \begin{tabular}{c|c|c}
        \ Single-site mean-field \ & \ Four-site mean-field \ & \ Numerical \ \\
        \hline
        -- & $\varepsilon_c^\mathrm{MF4} \simeq 4.00$ & $\varepsilon_c = 5.262(2)$
    \end{tabular}
    \caption{Comparison of the critical point activity obtained by four-site mean-field theory ($\varepsilon_c^\mathrm{MF4}$) and numerical simulations ($\varepsilon_c$).
    Single-site mean-field theory does not predict the critical point.}
    \label{tab_critical_point}
\end{table}

\begin{figure*}[t]
    \centering
    \includegraphics[scale=1]{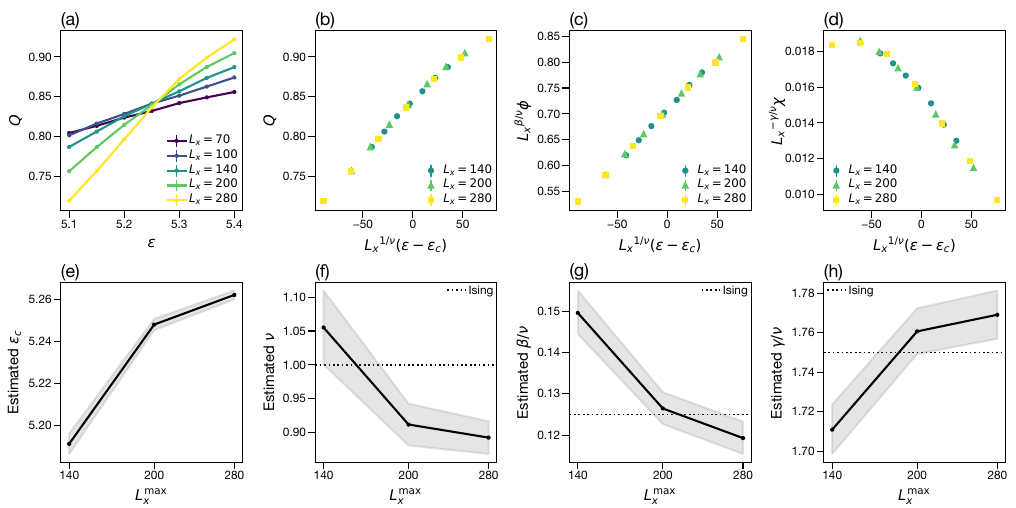}
    \caption{Finite-size scaling analysis for critical active phase separation in the symmetric limit ($\varepsilon_\mathrm{solute} = \varepsilon_\mathrm{solvent} = \varepsilon$).
    (a) Binder ratio $Q$ as a function of activity $\varepsilon$ for different system sizes with fixed $L_x / L_y = 2$.
    (b-d) Rescaled plots of the (b) Binder ratio $Q$, (c) order parameter $\phi$, and (d) susceptibility $\chi$, based on the finite-size scaling hypothesis [Eqs.~\eqref{eq_fss_Q}-\eqref{eq_fss_chi}] for the three largest system sizes.
    (e) Estimated critical point $\varepsilon_c$ as a function of $L_x^\mathrm{max}$, the maximum $L_x$ in the data set used for fitting.
    (f-h) Estimated critical exponents ($\nu$, $\beta / \nu$, and $\gamma / \nu$) as a function of $L_x^\mathrm{max}$, where the dotted lines suggest the Ising critical exponents.
    In (e-h), the estimation error is indicated by shading.}
    \label{fig_criticality}
\end{figure*}

\section{Criticality under bubble suppression}
\label{sec_criticality_under_bubble_supression}

In this section, we set $\rho = 0.5$ and consider the symmetric limit ($\varepsilon_\mathrm{solute} = \varepsilon_\mathrm{solvent} = \varepsilon$).
With this setup, the critical phase separation should appear on increasing activity $\varepsilon$, according to the symmetry of the model (Sec.~\ref{sec_lattice_model}) and four-site mean-field theory (Sec.~\ref{sec_four-site_mean-field_theory}).
Bubble formation has been thought to influence the critical behavior in active phase separation, potentially causing the critical exponents to deviate from the Ising universality class~\cite{Caballero2018, Shi2020, Speck2022, Cates2024}.
Since bubbles are suppressed under the symmetric limit (Sec.~\ref{sec_bubble_formation}), we expect to determine the critical exponents for active phase separation with negligible influence from bubbles.

We consider rectangular systems with a fixed aspect ratio ($L_x / L_y = 2$), where phase separation tends to appear along the $x$ axis.
Then, we define the order parameter for phase separation:
\begin{equation}
    \hat{\phi} := \frac{\sin (\pi / L_x)}{L_y} \left| \sum_{j, p} e^{-2 \pi i j_x / L_x} n_{j, \mathrm{solute}, p} \right|,
    \label{eq_order_parameter}
\end{equation}
where $j_x$ is the $x$ coordinate of site $j$, and $n_{j, \mathrm{solute}, p}$ is the local solute occupancy: $n_{j, \mathrm{solute}, p} = 1$ if site $j$ is occupied by a solute with polarity $p$ and $n_{j, \mathrm{solute}, p} = 0$ otherwise.
The prefactor is introduced so that $\hat{\phi} = 1$ for configurations that correspond to full phase separation along the $x$ axis.
Note that $\braket{\hat{\phi}^2}$ is proportional to the longest-wavelength component of the structure factor, $S (k_x = 2 \pi / L_x, k_y = 0)$, where $\braket{\cdots}$ is the ensemble average in the steady state.
Order parameters similar to Eq.~\eqref{eq_order_parameter} have been used in numerical studies of anisotropic phase separation in driven lattice gas models~\cite{Leung1991, Wang1996} and active particle models~\cite{Adachi2022, Nakano2024}.

To obtain the critical point $\varepsilon_c$ and critical exponents $(\nu, \beta, \gamma)$, we perform the finite-size scaling analysis~\cite{Landau2021} based on the scaling hypothesis: $\braket{\hat{\phi}^n} (\varepsilon, L_x) = {L_x}^{-n \beta / \nu} F_n \bm{(}{{L_x}^{1 / \nu}} (\varepsilon - \varepsilon_c) \bm{)}$ for $\varepsilon \simeq \varepsilon_c$ and $n \geq 1$, where $F_n$ is a scaling function.
We take the Binder ratio $Q := \braket{\hat{\phi}^2}^2 / \braket{\hat{\phi}^4}$, the (ensemble-averaged) order parameter $\phi := \braket{\hat{\phi}}$, and the susceptibility $\chi := L_x L_y (\braket{\hat{\phi}^2} - \braket{\hat{\phi}}^2)$ as observables.
According to the scaling hypothesis, these observables should follow
\begin{align}
    Q (\varepsilon, L_x) &= \tilde{Q} \bm{(}{{L_x}^{1 / \nu}} (\varepsilon - \varepsilon_c) \bm{)}, \label{eq_fss_Q} \\
    \phi (\varepsilon, L_x) &= {L_x}^{-\beta / \nu} \tilde{\phi} \bm{(}{{L_x}^{1 / \nu}} (\varepsilon - \varepsilon_c) \bm{)}, \\
    \chi (\varepsilon, L_x) &= {L_x}^{\gamma / \nu} \tilde{\chi} \bm{(}{{L_x}^{1 / \nu}} (\varepsilon - \varepsilon_c) \bm{)} \label{eq_fss_chi},
\end{align}
where $\tilde{Q}$, $\tilde{\phi}$, and $\tilde{\chi}$ are scaling functions.
Though the scaling hypothesis suggests the relation $2 \beta / \nu + \gamma / \nu = 2$, we estimate $(\nu, \beta / \nu, \gamma / \nu)$ as independent parameters from the data for $Q (\varepsilon, L_x)$, $\phi (\varepsilon, L_x)$, and $\chi (\varepsilon, L_x)$, which are obtained from steady-state configurations for several system sizes ($L_x = 70, 100, 140, 200, 280$ with $L_x / L_y = 2$) (see Appendix~\ref{app_statistical_properties} for details of sampling configurations).
If the critical exponents belong to the two-dimensional Ising universality class, we expect $\nu = \nu_\mathrm{Ising} := 1$, $\beta = \beta_\mathrm{Ising} := 0.125$, and $\gamma = \gamma_\mathrm{Ising} := 1.75$~\cite{Kadanoff1967}.

In Fig.~\ref{fig_criticality}(a), we plot the Binder ratio $Q (\varepsilon, L_x)$ as a function of $\varepsilon$ for different system sizes.
We find that $Q (\varepsilon, L_x)$ shows approximate crossing for $L_x \geq 100$.
The crossing point is expected to coincide with the critical point $\varepsilon_c$ since $Q (\varepsilon, L_x)$ should be independent of $L_x$ at $\varepsilon = \varepsilon_c$, i.e., $Q (\varepsilon = \varepsilon_c, L_x) = \tilde{Q} (0)$, according to Eq.~\eqref{eq_fss_Q}.

To determine $\varepsilon_c$ and $\nu$, we performed nonlinear least squares fitting of the data for $Q (\varepsilon, L_x)$ using a polynomial function.
Fitting was applied to each data set with three sequential system sizes (i.e., $L_x \in \{ 70, 100, 140 \}$, $\{ 100, 140, 200 \}$, or $\{ 140, 200, 280 \}$) to visualize higher-order corrections to the finite-size scaling hypothesis~\cite{Harada2015} (see Appendix~\ref{app_finite_size_scaling_analysis} for details).
Please refer to Appendix~\ref{app_validation_of_fitting_procedure} for validation of the same procedure within the equilibrium lattice gas model, which is known to show the Ising critical exponents~\cite{Kadanoff1967}.

In Figs.~\ref{fig_criticality}(e) and \ref{fig_criticality}(f), we plot the estimated values of $\varepsilon_c$ and $\nu$ along with their estimation errors (indicated by shading) as a function of $L_x^\mathrm{max}$, the maximum $L_x$ in the used data set.
For the data set with the largest $L_x^\mathrm{max}$ ($= 280$), the estimated exponent, $\nu = 0.89(2)$, is smaller than the Ising critical exponent, $\nu_\mathrm{Ising} = 1$, where the value in parentheses represents the error in the last significant digit.
In addition, the estimated critical point activity, $\varepsilon_c = 5.262(2)$, is larger than $\varepsilon_c^\mathrm{MF4} \simeq 4.00$ obtained by four-site mean-field theory (Sec.~\ref{sec_four-site_mean-field_theory}) as summarized in Table~\ref{tab_critical_point}, suggesting that long-range correlations beyond the four-site level make the homogeneous state more robust against activity.

We applied similar fitting procedures to the data for the order parameter $\phi (\varepsilon, L_x)$ and susceptibility $\chi (\varepsilon, L_x)$ to estimate $\beta / \nu$ and $\gamma / \nu$, respectively (see Appendix~\ref{app_finite_size_scaling_analysis} for details).
In Figs.~\ref{fig_criticality}(g) and \ref{fig_criticality}(h), we plot the obtained values of $\beta / \nu$ and $\gamma / \nu$ with their estimation errors as a function of $L_x^\mathrm{max}$.
For the data set with the largest $L_x^\mathrm{max}$ ($= 280$), the estimated values [$\beta / \nu = 0.119(4)$ and $\gamma / \nu = 1.77(1)$] are reasonably close to the Ising counterparts ($\beta_\mathrm{Ising} / \nu_\mathrm{Ising} = 0.125$ and $\gamma_\mathrm{Ising} / \nu_\mathrm{Ising} = 1.75$).
With the obtained critical exponents, we present the rescaled plots for $Q$, $\phi$, and $\chi$ in Figs.~\ref{fig_criticality}(b)-\ref{fig_criticality}(d).
Each plot suggests that the curves for different system sizes follow the universal scaling form, as expected from the scaling hypothesis [Eqs.~\eqref{eq_fss_Q}-\eqref{eq_fss_chi}].

The estimated $\varepsilon_c$ is not converged yet as a function of the system size, and the estimated exponent $\nu = 0.89(2)$ is smaller than the Ising counterpart $\nu_\mathrm{Ising} = 1$, while the two estimated ratios of exponents [$\beta / \nu = 0.119(4)$ and $\gamma / \nu = 1.77(1)$] are reasonably close to the Ising counterparts ($\beta_\mathrm{Ising} / \nu_\mathrm{Ising} = 0.125$ and $\gamma_\mathrm{Ising} / \nu_\mathrm{Ising} = 1.75$).
To conclude whether the asymptotic critical exponents for the infinite size limit will coincide with or deviate from the Ising values, we need larger-scale simulations.

\section{Summary and outlook}

In this paper, we have proposed an active binary mixture model, where we have found that bubble formation in active phase separation is controlled by the asymmetry in activities of solute and solvent ($\varepsilon_\mathrm{solute}$ and $\varepsilon_\mathrm{solvent}$).
With moderate activity asymmetry ($\varepsilon_\mathrm{solvent} \sim \varepsilon_\mathrm{solute} / 2$), we have found that the power-law scaling of the bubble size distribution is accessible in relatively small systems, compared to the active lattice gas limit ($\varepsilon_\mathrm{solvent} = 0$).
By employing mean-field theory that reproduces essential phase behaviors, we have discussed the potential mechanism for the bubble formation enhanced by activity asymmetry, as well as the competition of solute and solvent activities.
In the symmetric limit ($\varepsilon_\mathrm{solvent} = \varepsilon_\mathrm{solute}$), where bubbles are suppressed, we have conducted the finite-size scaling analysis for critical active phase separation, though larger-scale simulations are necessary to determine the universality class.
The mean-field analysis of our model and consideration based on AMB+ have suggested the following mechanism of the enhancement and suppression of bubbles: moderate activity asymmetry helps solvent bubbles grow in the solute-rich phase, while symmetric activity leads to macroscopic symmetry between the solute-rich and solvent-rich phases, which can result in macroscopic phase separation without bubble growth.

Bubbles have been suspected to change the critical exponents for active phase separation from the Ising critical exponents~\cite{Caballero2018, Shi2020, Speck2022, Cates2024}.
By enhancing the bubble formation within our model, we can potentially investigate this effect using relatively small system sizes, thereby reducing computational cost.
In addition, the mean-field dynamics [Eq.~\eqref{eq_mean_field_1site}], supplemented with noise, may provide another framework for examining the critical dynamics with bubble formation.

The controllability of bubbles by activity asymmetry may be observed in other active particle systems, such as a binary mixture of active Brownian particles~\cite{Kolb2020}.
More broadly, introducing solvent activity in active particle models will be interesting to study active matter phases in terms of symmetry between dense and dilute phases, given that phase separation appears ubiquitously~\cite{Chate2020}, e.g., in flocking~\cite{Solon2013, Solon2015aim, Solon2015} or active nematics~\cite{Ngo2014}.

\begin{acknowledgments}
We thank Kyogo Kawaguchi and Hiroyoshi Nakano for their continuous discussions and insightful comments.
We also thank Hugues Chat{\'e} and Kiyoshi Kanazawa for invaluable comments.
We acknowledge support by the RIKEN Information systems division for the use of the Supercomputer HOKUSAI BigWaterfall2.
This work was supported by JSPS KAKENHI Grant Number JP20K14435.
\end{acknowledgments}

\appendix

\section{Numerical simulations}

\subsection{Implementation of stochastic dynamics}
\label{app_implementation_of_stochastic_dynamics}

\begin{figure}[t]
    \centering
    \includegraphics[scale=1]{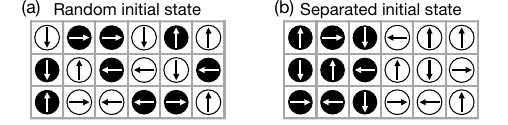}
    \caption{Two types of initial configurations.
    (a) Example of a random initial state.
    (b) Example of a separated initial state.
    The color and arrow indicate the species and polarity, respectively, in the same way as used in Fig.~\ref{fig_model}.}
    \label{appfig_initial_config}
\end{figure}

We consider two types of initial configurations: (a) ``random initial state'', in which both position and polarity of each particle are randomly chosen [Fig.~\ref{appfig_initial_config}(a)], and (b) ``separated initial state'', in which solutes and solvents are fully segregated along the $x$ axis, but polarity of each particle is randomly chosen [Fig.~\ref{appfig_initial_config}(b)].
The random initial states were used to examine critical phenomena (Fig.~\ref{fig_criticality}), while the separated initial states were used to examine the phase diagram (Fig.~\ref{fig_phase_diagram}) and bubble formation (Fig.~\ref{fig_bubble}).

To implement the stochastic dynamics of the lattice model (Fig.~\ref{fig_model}), we discretize the time with interval $\Delta t$.
In a single Monte Carlo (MC) step, we update the configuration as follows:
\begin{enumerate}[label=(\arabic*)]
    \item We randomly choose a particle, regardless of its species (i.e., solute or solvent).
    \item The chosen particle (of species $s$ with polarity $p$ at site $i$) (i) rotates its polarity or (ii) exchanges its position with one of the neighboring particles with the following probabilities.
    \begin{enumerate}[label=(\roman*)]
        \item The polarity $p$ rotates by $+90^\circ$ or $-90^\circ$, each with probability $\gamma \Delta t$.
        \item For each direction $l \in \{ +\hat{x}, +\hat{y}, -\hat{x}, -\hat{y} \}$, if the species $s$ is distinct from the species $s'$ of the particle at site $i + l$ (with polarity $p'$), the positions of these two particles are exchanged with probability $(1 + \varepsilon_s \delta_{p, l} + \varepsilon_{s'} \delta_{p', -l}) \Delta t / 2$.
        Here, $\hat{x}$ and $\hat{y}$ are the unit translations in the $x$ and $y$ directions, respectively, and the polarity indices $(\rightarrow, \uparrow, \leftarrow, \downarrow)$ are identified as $(+\hat{x}, +\hat{y}, -\hat{x}, -\hat{y})$, and $\delta_{a, b}$ is the Kronecker delta.
        Note that the factor $1/2$ appears in the exchange probability to compensate for the double counting of the same trial when selecting the particle at $i + l$ in procedure (1).
    \end{enumerate}
    \item We repeat procedures (1) and (2) $L _x L_y$ times and increment the time by $\Delta t$.
\end{enumerate}
To minimize the rejection probability in each trial, we set
\begin{equation}
    \Delta t = \frac{1}{(4 + \varepsilon_\mathrm{solute} + \varepsilon_\mathrm{solvent} + 3 \max \{ \varepsilon_\mathrm{solute}, \varepsilon_\mathrm{solvent} \})/2 + 2 \gamma}
    \label{appeq_timestep}
\end{equation}
in all the simulations conducted in this study.

See Ref.~\cite{code} for sample simulation codes, which can be executed on Google Colab to generate simulation videos.

\begin{figure}[t]
    \centering
    \includegraphics[scale=1]{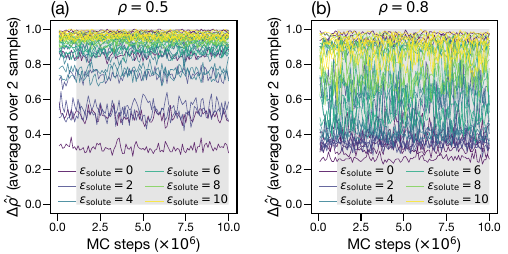}
    \caption{Time evolution of $\Delta \hat{\rho}'$.
    (a, b) Time-dependent $\Delta \hat{\rho}'$ for different sets of $(\varepsilon_\mathrm{solute}, \varepsilon_\mathrm{solvent})$, where the brightness suggests the solute activity, at (a) $\rho = 0.5$ or (b) $\rho = 0.8$.
    In (a) and (b), the gray area indicates the time points used for averaging to produce Figs.~\ref{fig_phase_diagram}(a) and \ref{fig_phase_diagram}(b), respectively.}
    \label{appfig_drho_timedep}
\end{figure}

\begin{figure*}[t]
    \centering
    \includegraphics[scale=1]{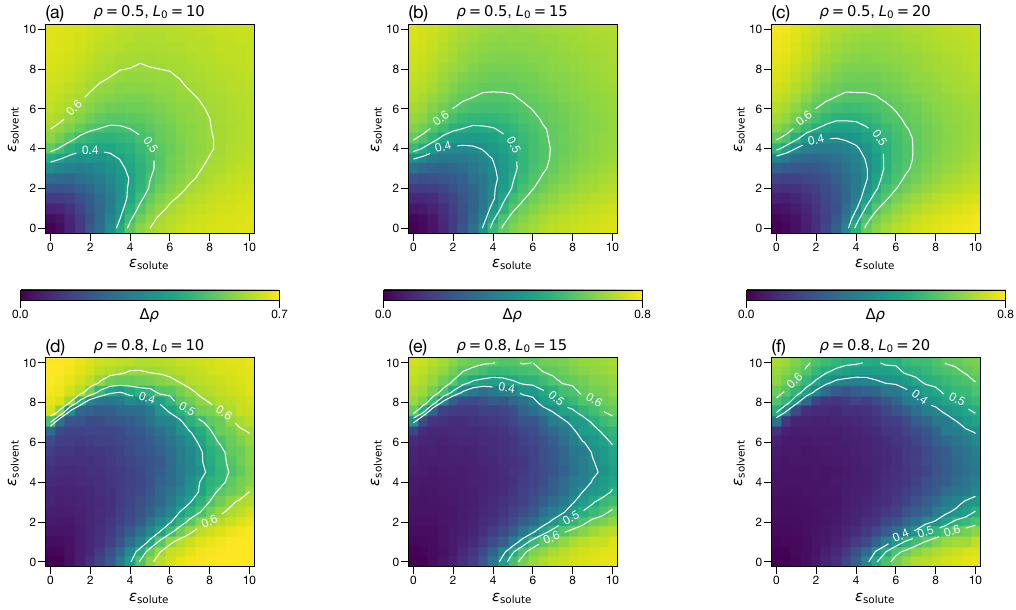}
    \caption{Weak dependence of the qualitative phase diagram on the definition of $\Delta \rho$.
    (a-c) Heatmaps and contour lines of the shifted density difference $\Delta \rho$, a degree of phase separation, for $\rho = 0.5$, obtained using subsystems of size $L_0 \times L_0$ with (a) $L_0 = 10$, (b) $L_0 = 15$, and (c) $L_0 = 20$.
    (d-f) Counterparts for $\rho = 0.8$.
    For all plots, the used system size is $(L_x, L_y) = (100, 50)$.
    Panels (a) and (d) correspond to Figs.~\ref{fig_phase_diagram}(a) and \ref{fig_phase_diagram}(b), respectively.}
    \label{appfig_phase_diagram_subsize}
\end{figure*}

\subsection{Qualitative phase diagram}
\label{app_numerical_qualitative_phase_diagram}

We define the most dense and dilute regions as the square subsystems (of size $L_0 \times L_0$) that contain the maximum and minimum number of solutes, respectively.
For a given configuration at certain activities $(\varepsilon_\mathrm{solute}, \varepsilon_\mathrm{solvent})$, we define $\Delta \hat{\rho}'$ as the difference in solute density between the most dense and dilute regions.
Writing the ensemble average of $\Delta \hat{\rho}'$ as $\Delta \rho'$, we define the shifted density difference as
\begin{equation}
    \Delta \rho := \Delta \rho' - \Delta \rho'_0,
\end{equation}
where $\Delta \rho'_0$ is the value of $\Delta \rho'$ at $\varepsilon_\mathrm{solute} = \varepsilon_\mathrm{solvent} = 0$ (i.e., baseline with no spatial correlation).

To plot the qualitative phase diagram for systems with $(L_x, L_y) = (100, 50)$ (Fig.~\ref{fig_phase_diagram}), we used $L_0 = 10$ and obtained $\Delta \rho'$ as the average of $\Delta \hat{\rho}'$ over two independent samples with 90 time points each, taken every $10^5$ MC steps after relaxation with $10^6$ MC steps using the separated initial state [Fig.~\ref{appfig_initial_config}(b)].
The time evolution of $\Delta \hat{\rho}'$ averaged over two independent samples for several values of $(\varepsilon_\mathrm{solute}, \varepsilon_\mathrm{solvent})$ is plotted in Fig.~\ref{appfig_drho_timedep}, where the gray area indicates the time points used for averaging.

Since the value of $\Delta \rho$ depends on $L_0$, we checked how the constant-$\Delta \rho$ lines depend on $L_0$.
In Fig.~\ref{appfig_phase_diagram_subsize}, we show the change in the heatmap of $\Delta \rho$ when $L_0$ increases as $L_0 = 10, 15, 20$, which suggests that the qualitative feature of the constant-$\Delta \rho$ lines is similar regardless of the specific value of $L_0$.

\begin{figure}[t]
    \centering
    \includegraphics[scale=1]{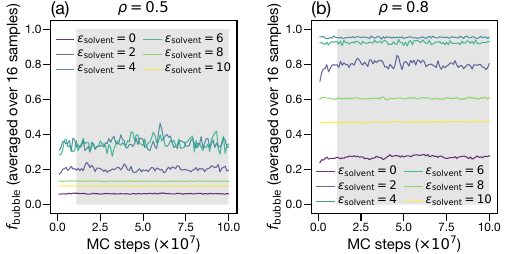}
    \caption{Time evolution of $f_\mathrm{bubble}$.
    (a, b) Time-dependent $f_\mathrm{bubble}$ for $(L_x, L_y) = (400, 200)$ and several $\varepsilon_\mathrm{solvent}$ at (a) $\rho = 0.5$ or (b) $\rho = 0.8$.
    In (a) and (b), the gray area indicates the time points used for averaging to produce Figs.~\ref{fig_bubble}(a-d) and \ref{fig_bubble}(e-h), respectively.}
    \label{appfig_fbubble_timedep}
\end{figure}

\begin{figure*}[t]
    \centering
    \includegraphics[scale=1]{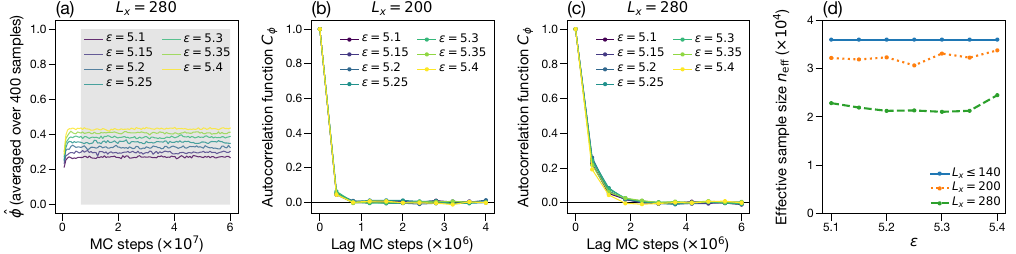}
    \caption{Statistical properties of order parameter $\hat{\phi}$ at $\rho = 0.5$ in the symmetric limit ($\varepsilon_\mathrm{solute} = \varepsilon_\mathrm{solvent} = \varepsilon$) for rectangular systems ($L_x / L_y = 2$).
    (a) Time evolution of $\hat{\phi}$ (averaged over independent samples) for different $\varepsilon$ with $L_x = 280$.
    The gray area indicates the time points used for averaging to produce Fig.~\ref{fig_criticality}.
    (b, c) Autocorrelation function $C_\phi$ as a function of time lag for different $\varepsilon$ and (b) $L_x = 200$ or (c) $L_x = 280$.
    (d) Effective sample size $n_\mathrm{eff}$ as a function of $\varepsilon$ for $L_x \leq 140$ (blue solid line), $L_x = 200$ (orange dotted line), and $L_x = 280$ (green dahsed line).}
    \label{appfig_phi_timedep_critical}
\end{figure*}

\begin{figure*}[t]
    \centering
    \includegraphics[scale=1]{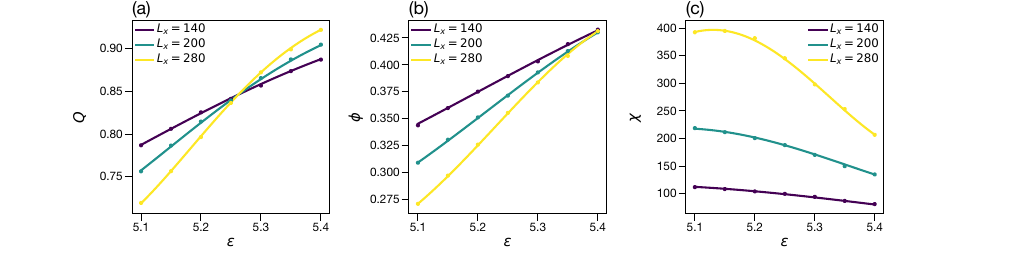}
    \caption{Fitting physical quantities to polynomial functions under the finite-size scaling hypothesis.
    (a-c) The fitted curves (colored solid lines) for the three largest system sizes ($L_x = 140, 200, 280$), shown with the observed data (colored dots) of the (a) Binder ratio $Q$, (b) order parameter $\phi$, and (c) susceptibility $\chi$.}
    \label{appfig_fitting_critical}
\end{figure*}

\subsection{Bubble formation}
\label{app_bubble_formation}

For each configuration, we define the bubbles as clusters of solvents, excluding the largest cluster, which is regarded as the bulk solvent-rich phase.
The bubble clusters are obtained using a Python package (scipy.ndimage.label~\cite{Virtanen2020}) with the default connectivity parameter (i.e., the nearest neighbors regarded as connected).

To calculate the bubble size distribution $N_\mathrm{bubble} (A)$ and bubble fraction $f_\mathrm{bubble}$ for each system size [$(L_x, L_y) = (100, 50), (200, 100), (400, 200)$] and solvent activity $\varepsilon_\mathrm{solvent}$, which are plotted in Fig.~\ref{fig_bubble}, we used configurations of 16 independent samples with 90 time points each, taken every $10^6$ MC steps after relaxation with $10^7$ MC steps using the separated initial state [Fig.~\ref{appfig_initial_config}(b)].
In Fig.~\ref{appfig_fbubble_timedep}, we show the time evolution of $f_\mathrm{bubble}$ obtained by averaging over 16 independent samples for the largest system size [$(L_x, L_y) = (400, 200)$] at (a) $\rho = 0.5$ or (b) $\rho = 0.8$.

\subsection{Critical phenomena in symmetric limit}

\subsubsection{Statistical properties of order parameter and related quantities}
\label{app_statistical_properties}

To obtain the ensemble-averaged quantities (i.e., Binder ratio $Q$, order parameter $\phi$, and susceptibility $\chi$) for each system size [$(L_x, L_y) = (70, 35), (100, 50), (140, 70), (200, 100), (280, 140)$] and activity $\varepsilon$, which are used in Fig.~\ref{fig_criticality}, we took configurations of 400 independent samples with 90 time points each, using the random initial state [Fig.~\ref{appfig_initial_config}(a)].
The samples are taken every $10^5$ MC steps after relaxation with $10^6$ MC steps for $(L_x, L_y) = (70, 35)$ and $(L_x, L_y) = (100, 50)$; taken every $4 \times 10^5$ MC steps after relaxation with $4 \times 10^6$ MC steps for $(L_x, L_y) = (140, 70)$ and $(L_x, L_y) = (200, 100)$; and taken every $6 \times 10^5$ MC steps after relaxation with $6 \times 10^6$ MC steps for $(L_x, L_y) = (280, 140)$.
In Fig.~\ref{appfig_phi_timedep_critical}(a), we plot the time evolution of the order parameter $\hat{\phi}$ averaged over 400 independent samples for the largest system size [$(L_x, L_y) = (280, 140)$].

We calculated the effective sample size $n_\mathrm{eff}$ in the following way.
The normalized autocorrelation function $C_\phi (t_\mathrm{lag})$ is defined as~\cite{Landau2021}
\begin{equation}
    C_\phi (t_\mathrm{lag}) := \frac{\braket{\delta \hat{\phi}(t) \, \delta \hat{\phi}(t + t_\mathrm{lag})}}{\braket{(\delta \hat{\phi})^2}},
\end{equation}
where $t_\mathrm{lag}$ is the time lag, $\delta \hat{\phi} (t) := \hat{\phi} (t) - \braket{\hat{\phi}}$ is the order parameter fluctuation for a configuration at time $t$, and $\braket{\cdots}$ means the ensemble average in the steady state.
Practically, we calculated $C_\phi (t_\mathrm{lag})$ by averaging over independent samples and time points after relaxation [see the gray area in Fig.~\ref{appfig_phi_timedep_critical}(a) for $L_x = 280$].
For $L_x = 70, 100, 140$, we confirmed that $C_\phi$ is small ($C_\phi < 0.015$) at the time lag between two consecutive observation points; thus, the effective sample size is calculated as $n_\mathrm{eff} = n_\mathrm{sample} n_\mathrm{time}$, where $n_\mathrm{sample} = 400$ and $n_\mathrm{time} = 90$ are the total numbers of independent samples and observation time points, respectively.
On the other hand, for $L_x = 200$ and $L_x = 280$, we found non-negligible autocorrelation as shown in Figs.~\ref{appfig_phi_timedep_critical}(b) and \ref{appfig_phi_timedep_critical}(c), respectively; thus, the effective sample size is calculated as $n_\mathrm{eff} = n_\mathrm{sample} n_\mathrm{time} / [1 + 2 \sum_{1 \leq i \leq 10} C(t_{\mathrm{lag}, i})]$ by taking into account the correction by autocorrelation~\cite{Landau2021}, where $t_{\mathrm{lag}, i}$ is the $i$th smallest time lag among observation time points.
The calculated $n_\mathrm{eff}$ is plotted in Fig.~\ref{appfig_phi_timedep_critical}(d), which shows the decrease in the effective sample size for $L_x = 200$ and $L_x = 280$.

The standard error for $Q$, $\phi$, and $\chi$ are obtained as follows.
First, $\sigma_\phi$ (standard deviation for $\phi$) is obtained from the configurations used to calculate the mean value $\phi$.
Then, using the error propagation formula, we obtain $\sigma_Q$ and $\sigma_\chi$ (standard deviations for $Q$ and $\chi$, respectively).
Lastly, using the effective sample size $n_\mathrm{eff}$ obtained above, we estimate the standard errors as $\sigma_q / \sqrt{n_\mathrm{eff}}$ for $q = Q, \phi, \chi$.
These standard errors are used for error bars in Figs.~\ref{fig_criticality}(a)-\ref{fig_criticality}(d).

\begin{figure*}[t]
    \centering
    \includegraphics[scale=1]{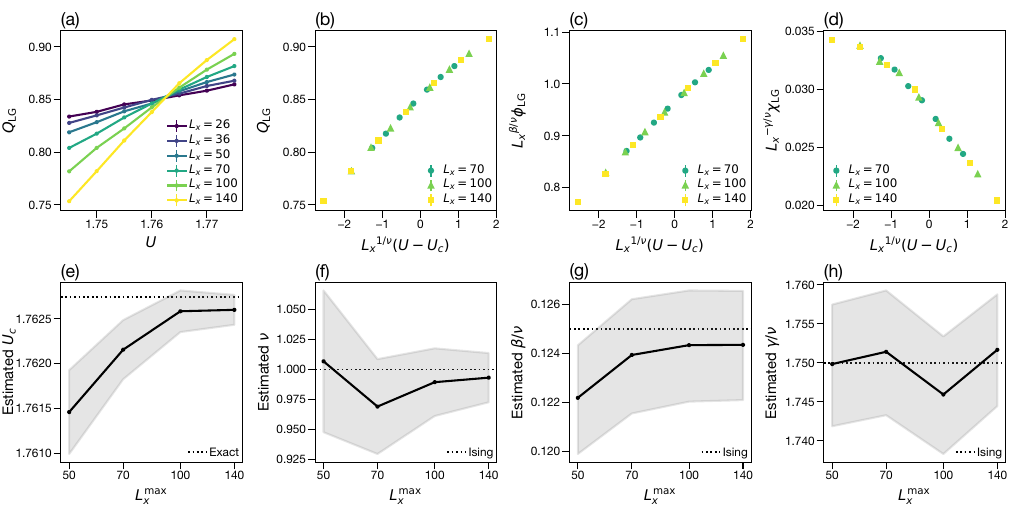}
    \caption{Ising universality for critical phenomena in the equilibrium lattice gas model.
    (a) Binder ratio $Q_\mathrm{LG}$ as a function of interaction strength $U$ for different system sizes with fixed $L_x / L_y = 2$.
    (b-d) Rescaled plots of the (b) Binder ratio $Q_\mathrm{LG}$, (c) order parameter $\phi_\mathrm{LG}$, and (d) susceptibility $\chi_\mathrm{LG}$, based on the finite-size scaling hypothesis for the three largest system sizes.
    (e) Estimated critical point $U_c$ as a function of $L_x^\mathrm{max}$, the maximum $L_x$ in the data set used for fitting, where the dotted line suggests the exact critical point $U_c^\mathrm{exact} = 2 \ln (1 + \sqrt{2})$~\cite{Onsager1944}.
    (f-h) Estimated critical exponents ($\nu$, $\beta / \nu$, and $\gamma / \nu$) as a function of $L_x^\mathrm{max}$, where the dotted lines suggest the Ising critical exponents.
    In (e-h), the estimation error is indicated by shading.}
    \label{appfig_criticality_mlg}
\end{figure*}

\subsubsection{Finite-size scaling analysis}
\label{app_finite_size_scaling_analysis}

To determine $\varepsilon_c$ and $\nu$, we performed nonlinear least squares fitting of the data for $Q (\varepsilon, L_x)$ using a polynomial function that satisfies the scaling hypothesis [Eq.~\eqref{eq_fss_Q}].
The fitting was carried out with a Python package (scipy.optimize.curve{\_}fit~\cite{Virtanen2020}).
Specifically, we used $a_0 + a_1 {L_x}^{1 / \tilde{\nu}} (\varepsilon - \tilde{\varepsilon}_c) + a_2 {L_x}^{2 / \tilde{\nu}} (\varepsilon - \tilde{\varepsilon}_c)^2 + a_3 {L_x}^{3 / \tilde{\nu}} (\varepsilon - \tilde{\varepsilon}_c)^3 + a_4 {L_x}^{4 / \tilde{\nu}} (\varepsilon - \tilde{\varepsilon}_c)^4$ as the fitting function, where $\{ a_i \}_{i = 0}^4$, $\tilde{\varepsilon}_c$, and $\tilde{\nu}$ are fitting parameters.
We used the standard error of $Q$ (see Appendix~\ref{app_statistical_properties}) as uncertainty.
In Fig.~\ref{appfig_fitting_critical}(a), we show the fitted curves for the three largest system sizes ($L_x = 140, 200, 280$).
The optimal values, $\tilde{\varepsilon}_c = \varepsilon_c$ and $\tilde{\nu} = \nu$, which are obtained using the data set for three sequential system sizes, are plotted in Figs.~\ref{fig_criticality}(e) and \ref{fig_criticality}(f) with black dots, shown with fitting errors $\pm \delta \varepsilon_c$ and $\pm \delta \nu$ suggested by shading, respectively.

For each pair taken from $\{ \varepsilon_c$, $\varepsilon_c + \delta \varepsilon_c$, $\varepsilon_c - \delta \varepsilon_c \}$ and $\{ \nu, \nu + \delta \nu, \nu - \delta \nu \}$, which is used as a representative value of $(\tilde{\varepsilon}_c, \tilde{\nu})$, we performed fitting of the data for $\phi (\varepsilon, L_x)$.
Specifically, for each pair $(\tilde{\varepsilon}_c, \tilde{\nu})$ [e.g., $(\tilde{\varepsilon}_c, \tilde{\nu}) = (\varepsilon_c + \delta \varepsilon_c, \nu - \delta \nu)$], we used ${L_x}^{-\tilde{\beta}'} [b_0 + b_1 {L_x}^{1 / \tilde{\nu}} (\varepsilon - \tilde{\varepsilon}_c) + b_2 {L_x}^{2 / \tilde{\nu}} (\varepsilon - \tilde{\varepsilon}_c)^2 + b_3 {L_x}^{3 / \tilde{\nu}} (\varepsilon - \tilde{\varepsilon}_c)^3 + b_4 {L_x}^{4 / \tilde{\nu}} (\varepsilon - \tilde{\varepsilon}_c)^4]$ as the fitting function, where $\{ b_i \}_{i = 0}^4$ and $\tilde{\beta}'$ are fitting parameters.
We used the standard error of $\phi$ (see Appendix~\ref{app_statistical_properties}) as uncertainty.
In Fig.~\ref{appfig_fitting_critical}(b), we show the fitted curves for the three largest system sizes ($L_x = 140, 200, 280$) when taking $(\tilde{\varepsilon}_c, \tilde{\nu}) = (\varepsilon_c, \nu)$.
We obtained the optimal value, $\tilde{\beta}' = \beta'$ ($= \beta / \nu$), for each pair $(\tilde{\varepsilon}_c, \tilde{\nu})$; we chose $\beta'$ at $(\tilde{\varepsilon}_c, \tilde{\nu}) = (\varepsilon_c, \nu)$ as the estimated value, which is plotted in Fig.~\ref{fig_criticality}(g) with black dots.
We also obtained the fitting error $\pm \delta \beta'$ for each pair $(\tilde{\varepsilon}_c, \tilde{\nu})$; the maximum value of $\beta' + \delta \beta'$ over all (i.e., nine) pairs of $(\tilde{\varepsilon}_c, \tilde{\nu})$ was used for the upper estimation error, and the minimum value of $\beta' - \delta \beta'$ was used for the lower estimation error, as plotted in Fig.~\ref{fig_criticality}(g) with shading.

We obtained $\gamma'$ ($= \gamma / \nu$) and its estimation error by fitting the data for $\chi (\varepsilon, L_x)$ in the same way as used for $\beta'$ ($= \beta / \nu$).
Specifically, we used ${L_x}^{\tilde{\gamma}'} [c_0 + c_1 {L_x}^{1 / \tilde{\nu}} (\varepsilon - \tilde{\varepsilon}_c) + c_2 {L_x}^{2 / \tilde{\nu}} (\varepsilon - \tilde{\varepsilon}_c)^2 + c_3 {L_x}^{3 / \tilde{\nu}} (\varepsilon - \tilde{\varepsilon}_c)^3 + c_4 {L_x}^{4 / \tilde{\nu}} (\varepsilon - \tilde{\varepsilon}_c)^4]$ as the fitting function, where $\{ c_i \}_{i = 0}^4$ and $\tilde{\gamma}'$ are fitting parameters, and used the standard error of $\chi$ (see Appendix~\ref{app_statistical_properties}) as uncertainty.
In Fig.~\ref{appfig_fitting_critical}(c), we show the fitted curves for the three largest system sizes ($L_x = 140, 200, 280$) when taking $(\tilde{\varepsilon}_c, \tilde{\nu}) = (\varepsilon_c, \nu)$.
The estimated $\gamma'$ and its error are plotted in Fig.~\ref{fig_criticality}(h).

\subsubsection{Validation of fitting procedure for equilibrium lattice gas}
\label{app_validation_of_fitting_procedure}

To confirm the validity of the fitting procedure explained in Appendix~\ref{app_finite_size_scaling_analysis} for a well-known model, we consider equilibrium lattice gas, where $N$ ($= \rho_\mathrm{LG} L_x L_y$) particles are interacting in equilibrium on a rectangular lattice of size $(L_x, L_y)$ with periodic boundary conditions along the $x$ and $y$ axes.
We write the equilibrium distribution for this model as $P_\mathrm{eq}(\bm{n})$, where $\bm{n} := \{ n_i \}$ is the particle configuration with $n_i \in \{ 0, 1 \}$ indicating the occupancy at site $i$.
We assume $\ln P_\mathrm{eq}(\bm{n}) = U \sum_{\braket{i, j}} n_i n_j + \mathrm{const.}$, where $U$ is the dimensionless attractive interaction strength, and $\sum_{\braket{i, j}} (\cdots)$ represents the summation over nearest-neighbor sites.
It has been known that, in the thermodynamic limit, this model undergoes critical phase separation at $U = U_c^\mathrm{exact} := 2 \ln (1 + \sqrt{2})$ and $\rho_\mathrm{LG} = 0.5$ with critical exponents in the two-dimensional Ising universality class~\cite{Onsager1944}.

We implement the stochastic dynamics in which the steady-state distribution coincides with $P_\mathrm{eq}(\bm{n})$ as follows.
In a single MC step, we update the configuration as follows:
\begin{enumerate}[label=(\arabic*)]
    \item We randomly choose a particle and an empty site.
    \item The chosen particle hops to the chosen empty site with probability $1 / [1 + \exp(-U \Delta n)]$, where $\Delta n$ is the change in the total number of particles adjacent to the chosen particle as a result of the hop.
    \item We repeat procedures (1) and (2) $N$ times.
\end{enumerate}
We stress that we allow not only the nearest-neighbor hopping, which is used in the Kawasaki dynamics, but also any nonlocal hopping, which accelerates the convergence to the steady state~\cite{Landau2021, Tamayo1989, Bray1991, Tamayo1991, Moseley1992}.
Note that the nonequilibrium extension of similar models with nonlocal hopping has been numerically found to undergo critical phase separation with the Ising critical exponents~\cite{Zakine2024}.

To apply the finite-size scaling analysis, we consider different system sizes ($L_x = 26, 36, 50, 70, 100, 140$) with fixed $L_x / L_y = 2$ and several values of $U$ near $U_c^\mathrm{exact}$ at $\rho_\mathrm{LG} = 0.5$.
We define the order parameter $\hat{\phi}_\mathrm{LG}$ in the same way as used in Eq.~\eqref{eq_order_parameter}:
\begin{equation}
    \hat{\phi}_\mathrm{LG} := \frac{\sin (\pi / L_x)}{L_y} \left| \sum_j e^{-2 \pi i j_x / L_x} n_j \right|.
\end{equation}
To obtain the Binder ratio $Q_\mathrm{LG} := \braket{{\hat{\phi}_\mathrm{LG}}^2}^2 / \braket{\hat{\phi}_\mathrm{LG}}^4$, (ensemble-averaged) order parameter $\phi_\mathrm{LG} := \braket{\hat{\phi}_\mathrm{LG}}$, and susceptibility $\chi_\mathrm{LG} := L_x L_y (\braket{{\hat{\phi}_\mathrm{LG}}^2} - \braket{\hat{\phi}_\mathrm{LG}}^2)$, we took configurations of 1000 independent samples with 90 time points each, using the initial state with a random distribution of particles.
The samples are taken every $10^3$ MC steps after relaxation with $10^4$ MC steps for $L_x = 26$ and $L_x = 36$; taken every $10^4$ MC steps after relaxation with $10^5$ MC steps for $L_x = 50$, $L_x = 70$, and $L_x = 100$; and taken every $5 \times 10^4$ MC steps after relaxation with $5 \times 10^5$ MC steps for $L_x = 140$.
We confirmed that normalized autocorrelation function for $\hat{\phi}_\mathrm{LG}$ is small ($< 0.01$) at the time lag between two consecutive observation points for all $L_x$ and $U$; thus, the effective sample size is calculated as $n_\mathrm{eff} = 1000 \times 90 = 9 \times 10^4$ for each $L_x$ and $U$.

In the same way as used in Appendix~\ref{app_finite_size_scaling_analysis}, we performed fitting of the obtained data based on the finite-size scaling hypothesis.
The results are shown in Fig.~\ref{appfig_criticality_mlg}.
First, Fig.~\ref{appfig_criticality_mlg}(a) shows approximate crossing of the Binder ratio $Q_\mathrm{LG}$, suggesting the value of the critical point $U_c$.
In Fig.~\ref{appfig_criticality_mlg}(e), we plot the estimated $U_c$ (black dots) as a function of $L_x^\mathrm{max}$, the maximum $L_x$ in the used data set with three sequential system sizes (e.g., $L_x^\mathrm{max} = 100$ for $L_x \in \{ 50, 70, 100 \}$).
This suggests that the fitting procedure works well to obtain $U_c$ close to $U_c^\mathrm{exact}$ (dotted line).
Then, we present the estimated critical exponents ($\nu$, $\beta / \nu$, and $\gamma / \nu$) as a function of $L_x$ in Figs.~\ref{appfig_criticality_mlg}(f)-\ref{appfig_criticality_mlg}(h) with black dots, showing that the estimated values are close to the Ising exponents (dotted lines), as expected from theory~\cite{Onsager1944}.
Each of the rescaled plots shown in Figs.~\ref{appfig_criticality_mlg}(b)-\ref{appfig_criticality_mlg}(d) suggests the universal scaling form independent of $L_x$, consistent with the finite-size scaling hypothesis.
Quantitatively, for the data set with the largest $L_x^\mathrm{max}$, the fitting procedure resulted in $U_c = 1.7626(2)$, $\nu = 0.99(2)$, $\beta / \nu = 0.124(2)$, and $\gamma / \nu = 1.752(7)$, which are close to the exact values, $U_c^\mathrm{exact} = 1.7627...$, $\nu_\mathrm{Ising} = 1$, $\beta_\mathrm{Ising} / \nu_\mathrm{Ising} = 0.125$, and $\gamma_\mathrm{Ising} / \nu_\mathrm{Ising} = 1.75$, respectively.

\begin{figure}[t]
    \centering
    \includegraphics[scale=1]{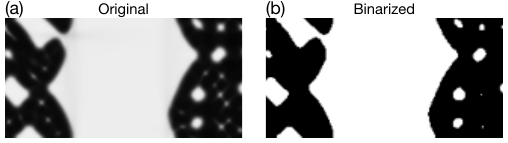}
    \caption{Binarized solute density in mean-field dynamics.
    (a, b) Example of snapshots for (a) the original solute density $\rho_{i, \mathrm{solute}}$ and (b) the binarized counterpart.
    The used parameters are $(L_x, L_y) = (200, 100)$, $\rho = 0.5$, and $(\varepsilon_\mathrm{solute}, \varepsilon_\mathrm{solvent}) = (10, 3)$.}
    \label{appfig_mean-field_binarize}
\end{figure}

\begin{figure}[t]
    \centering
    \includegraphics[scale=1]{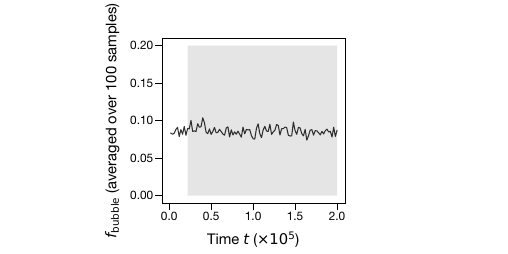}
    \caption{Time evolution of $f_\mathrm{bubble}$ in mean-field dynamics.
    The used parameters are $(L_x, L_y) = (200, 100)$, $\rho = 0.5$, and $(\varepsilon_\mathrm{solute}, \varepsilon_\mathrm{solvent}) = (10, 3)$.
    The gray area indicates the time points used for averaging to produce Figs.~\ref{fig_mean-field_bubble}(d) and \ref{fig_mean-field_bubble}(e).}
    \label{appfig_mean-field_fbubble_timedep}
\end{figure}

\section{Mean-field theory}

\subsection{Bubble formation in mean-field dynamics}
\label{app_bubble_formation_in_mean-field_dynamics}

We specify the solvent bubbles for the mean-field dynamics as follows.
First, we binarize the local solute density $\rho_{i, \mathrm{solute}}$ with a threshold value $0.5$ (see Fig.~\ref{appfig_mean-field_binarize} for an example).
Then, in the same way as used for the stochastic dynamics (see Appendix~\ref{app_bubble_formation}), we define the bubbles as clusters of solvents, excluding the largest cluster, which is regarded as the bulk solvent-rich phase.

To calculate the bubble size distribution $N_\mathrm{bubble} (A)$ and bubble fraction $f_\mathrm{bubble}$ at $\rho = 0.5$ and $(\varepsilon_\mathrm{solute}, \varepsilon_\mathrm{solvent}) = (10, 3)$ for each system size [$(L_x, L_y) = (100, 50), (140, 70), (200, 100)$], which are plotted in Figs.~\ref{fig_mean-field_bubble}(d) and \ref{fig_mean-field_bubble}(e), we used configurations of 100 independent samples with 90 time points each, taken every $10^5$ steps with time step $dt = 0.02$ after relaxation with $10^6$ steps using the separated initial state [Fig.~\ref{appfig_initial_config}(b)].
In Fig.~\ref{appfig_mean-field_fbubble_timedep}, we show the time evolution of $f_\mathrm{bubble}$ obtained by averaging over 100 independent samples for the largest system size [$(L_x, L_y) = (200, 100)$].

\begin{widetext}

\subsection{Linear stability analysis for single-site mean-field theory}
\label{app_linear_stability_analysis_for_single-site_mean-field_theory}

The explicit form of Eq.~\eqref{eq_mean_field_1site} is given as
\begin{align}
    \partial_t \rho_{i, s, p} & = \sum_{l \in \{ \pm \hat{x}, \pm \hat{y} \}} \sum_{p'} (\rho_{i, \bar{s}, p'} \rho_{i + l, s, p} - \rho_{i + l, \bar{s}, p'} \rho_{i, s, p}) + \sum_{p'} \varepsilon_s (\rho_{i, \bar{s}, p'} \rho_{i - p, s, p} - \rho_{i + p, \bar{s}, p'} \rho_{i, s, p}) + \sum_{p'} \varepsilon_{\bar{s}} (\rho_{i, \bar{s}, p'} \rho_{i + p', s, p} - \rho_{i - p', \bar{s}, p'} \rho_{i, s, p}) \nonumber \\
    & \quad + \gamma (\rho_{i, s, R p} + \rho_{i, s, R^{-1} p} - 2 \rho_{i, s, p}).
\end{align}
We linearize this equation in terms of the local fluctuations around the homogeneous state [i.e., $\rho_{i, \mathrm{solute}, p} = \rho / 4$ and $\rho_{i, \mathrm{solvent}, p} = (1 - \rho) / 4$ for any $i$ or $p$].
Using $\rho_{\bm{k}, s, p} = \sum_j e^{-i \bm{k} \cdot j} \rho_{j, s, p}$, we obtain
\begin{align}
    \partial_t \rho_{\bm{k}, s, p} & = -\frac{1}{4} \sum_{p'} 2 (2 - \cos k_x - \cos k_y) (\rho_{\bar{s}} \rho_{\bm{k}, s, p} - \rho_s \rho_{\bm{k}, \bar{s}, p'}) - \frac{1}{4} \sum_{p'} \varepsilon_s [(1 - e^{-i \bm{k} \cdot p}) \rho_{\bar{s}} \rho_{\bm{k}, s, p} - (1 - e^{i \bm{k} \cdot p}) \rho_s \rho_{\bm{k}, \bar{s}, p'}] \nonumber \\
    & \quad - \frac{1}{4} \sum_{p'} \varepsilon_{\bar{s}} [(1 - e^{i \bm{k} \cdot p'}) \rho_{\bar{s}} \rho_{\bm{k}, s, p} - (1 - e^{-i \bm{k} \cdot p'}) \rho_s \rho_{\bm{k}, \bar{s}, p'}] - \gamma (2 \rho_{\bm{k}, s, p} - \rho_{\bm{k}, s, R p} - \rho_{\bm{k}, s, R^{-1} p}) \nonumber \\
    & = -2 (2 - \cos k_x - \cos k_y) \left[ \rho_{\bar{s}} \rho_{\bm{k}, s, p} - \frac{1}{4} \rho_s \sum_{p'} \rho_{\bm{k}, \bar{s}, p'} \right] - \varepsilon_s \left[ (1 - e^{-i \bm{k} \cdot p}) \rho_{\bar{s}} \rho_{\bm{k}, s, p} - \frac{1}{4} (1 - e^{i \bm{k} \cdot p}) \rho_s \sum_{p'} \rho_{\bm{k}, \bar{s}, p'} \right] \nonumber \\
    & \quad - \varepsilon_{\bar{s}} \left[ \frac{1}{2} (2 - \cos k_x - \cos k_y) \rho_{\bar{s}} \rho_{\bm{k}, s, p} - \frac{1}{4} \rho_s \sum_{p'} (1 - e^{-i \bm{k} \cdot p'}) \rho_{\bm{k}, \bar{s}, p'} \right] - \gamma (2 \rho_{\bm{k}, s, p} - \rho_{\bm{k}, s, R p} - \rho_{\bm{k}, s, R^{-1} p}),
    \label{eq_mean_field_1site_linearized_explicit}
\end{align}
where $\rho_\mathrm{solute} := \rho$, $\rho_\mathrm{solvent} := 1 - \rho$, and $\bar{s}$ represents the species opposite to $s$.
Considering $\rho_{\bm{k}, s, p}$ as a component of $8$-dimensional vector $\bm{\rho}_{\bm{k}}$, we can write Eq.~\eqref{eq_mean_field_1site_linearized_explicit} in the form of $\partial_t \bm{\rho}_{\bm{k}} = R_{\bm{k}}^\mathrm{MF1} \bm{\rho}_{\bm{k}}$ with $8 \times 8$ matrix $R_{\bm{k}}^\mathrm{MF1}$.

To examine linear stability against small-wavenumber fluctuations, we expand the coefficients in Eq.~\eqref{eq_mean_field_1site_linearized_explicit} in terms of $\bm{k}$ up to $O(|\bm{k}|^4)$:
\begin{align}
    \partial_t \rho_{\bm{k}, s, p} & = -\left( |\bm{k}|^2 - \frac{1}{12} {k_x}^4 - \frac{1}{12} {k_y}^4 \right) \left[ \rho_{\bar{s}} \rho_{\bm{k}, s, p} - \frac{1}{4} \rho_s \sum_{p'} \rho_{\bm{k}, \bar{s}, p'} \right] - \varepsilon_s \left[ i \bm{k} \cdot p + \frac{1}{2} (\bm{k} \cdot p)^2 - \frac{i}{6} (\bm{k} \cdot p)^3 - \frac{1}{24} (\bm{k} \cdot p)^4 \right] \rho_{\bar{s}} \rho_{\bm{k}, s, p} \nonumber \\
    & \quad - \frac{1}{4} \varepsilon_s \left[ i \bm{k} \cdot p - \frac{1}{2} (\bm{k} \cdot p)^2 - \frac{i}{6} (\bm{k} \cdot p)^3 + \frac{1}{24} (\bm{k} \cdot p)^4 \right] \rho_s \sum_{p'} \rho_{\bm{k}, \bar{s}, p'} - \frac{1}{4} \varepsilon_{\bar{s}} \left( |\bm{k}|^2 - \frac{1}{12} {k_x}^4 - \frac{1}{12} {k_y}^4 \right) \rho_{\bar{s}} \rho_{\bm{k}, s, p} \nonumber \\
    &\quad + \frac{1}{4} \varepsilon_{\bar{s}} \rho_s \sum_{p'} \left[ i \bm{k} \cdot p' + \frac{1}{2} (\bm{k} \cdot p')^2 - \frac{i}{6} (\bm{k} \cdot p')^3 - \frac{1}{24} (\bm{k} \cdot p')^4 \right] \rho_{\bm{k}, \bar{s}, p'} - \gamma (2 \rho_{\bm{k}, s, p} - \rho_{\bm{k}, s, Rp} - \rho_{\bm{k}, s, R^{-1}p}).
    \label{eq_mean_field_1site_linearized}
\end{align}

We introduce the density fluctuation $\rho_{\bm{k}, s} := \sum_p \rho_{\bm{k}, s, p}$, polarity density fluctuation $\bm{w}_{\bm{k}, s} := \sum_p p \rho_{\bm{k}, s, p}$, and nematicity density fluctuation $\nu_{\bm{k}, s} := \sum_p ({p_x}^2 - {p_y}^2) \rho_{\bm{k}, s, p}$.
Noticing that $\rho_{\bm{k}, \bar{s}} = - \rho_{\bm{k}, s}$ and $\rho_{\bar{s}} = 1 - \rho_s$, we rewrite Eq.~\eqref{eq_mean_field_1site_linearized} as
\begin{align}
    \partial_t \rho_{\bm{k}, s} & = - \left( 1 + \frac{1}{4} \varepsilon_s + \frac{1}{4} \varepsilon_{\bar{s}} \right) \left( |\bm{k}|^2 - \frac{1}{12} {k_x}^4 - \frac{1}{12} {k_y}^4 \right) \rho_{\bm{k}, s} - i \varepsilon_s (1 - \rho_s) \left( \bm{k} \cdot \bm{w}_{\bm{k}, s} - \frac{1}{6} {k_x}^3 w_{x, \bm{k}, s} - \frac{1}{6} {k_y}^3 w_{y, \bm{k}, s} \right) \nonumber \\
    & \quad + i \varepsilon_{\bar{s}} \rho_s \left( \bm{k} \cdot \bm{w}_{\bm{k}, \bar{s}} - \frac{1}{6} {k_x}^3 w_{x, \bm{k}, \bar{s}} - \frac{1}{6} {k_y}^3 w_{y, \bm{k}, \bar{s}} \right) - \frac{1}{4} \varepsilon_s (1 - \rho_s) \left( {k_x}^2 - {k_y}^2 - \frac{1}{12} {k_x}^4 + \frac{1}{12} {k_y}^4 \right) \nu_{\bm{k}, s} \nonumber \\
    & \quad + \frac{1}{4} \varepsilon_{\bar{s}} \rho_s \left( {k_x}^2 - {k_y}^2 - \frac{1}{12} {k_x}^4 + \frac{1}{12} {k_y}^4 \right) \nu_{\bm{k}, \bar{s}}, \label{eq_mean_field_1site_linearized_rho} \\
    \partial_t w_{x, \bm{k}, s} & = -(1 - \rho_s) \left( |\bm{k}|^2 - \frac{1}{12} {k_x}^4 - \frac{1}{12} {k_y}^4 \right) w_{x, \bm{k}, s} - \frac{1}{2} \varepsilon_s (1 - \rho_s) \left[ i k_x (\rho_{\bm{k}, s} + \nu_{\bm{k}, s}) + {k_x}^2 w_{x, \bm{k}, s} - \frac{i}{6} {k_x}^3 (\rho_{\bm{k}, s} + \nu_{\bm{k}, s}) - \frac{1}{12} {k_x}^4 w_{x, \bm{k}, s} \right] \nonumber \\
    & \quad + \frac{1}{2} \varepsilon_s \rho_s \left( i k_x - \frac{i}{6} {k_x}^3 \right) \rho_{\bm{k}, s} - \frac{1}{4} \varepsilon_{\bar{s}} (1 - \rho_s) \left( |\bm{k}|^2 - \frac{1}{12} {k_x}^4 - \frac{1}{12} {k_y}^4 \right) w_{x, \bm{k}, s} - 2 \gamma w_{x, \bm{k}, s}, \label{eq_mean_field_1site_linearized_wx} \\
    \partial_t w_{y, \bm{k}, s} & = -(1 - \rho_s) \left( |\bm{k}|^2 - \frac{1}{12} {k_x}^4 - \frac{1}{12} {k_y}^4 \right) w_{y, \bm{k}, s} - \frac{1}{2} \varepsilon_s (1 - \rho_s) \left[ i k_y (\rho_{\bm{k}, s} - \nu_{\bm{k}, s}) + {k_y}^2 w_{y, \bm{k}, s} - \frac{i}{6} {k_y}^3 (\rho_{\bm{k}, s} - \nu_{\bm{k}, s}) - \frac{1}{12} {k_y}^4 w_{y, \bm{k}, s} \right] \nonumber \\
    & \quad + \frac{1}{2} \varepsilon_s \rho_s \left( i k_y - \frac{i}{6} {k_y}^3 \right) \rho_{\bm{k}, s} - \frac{1}{4} \varepsilon_{\bar{s}} (1 - \rho_s) \left( |\bm{k}|^2 - \frac{1}{12} {k_x}^4 - \frac{1}{12} {k_y}^4 \right) w_{y, \bm{k}, s} - 2 \gamma w_{y, \bm{k}, s}, \label{eq_mean_field_1site_linearized_wy} \\
    \partial_t \nu_{\bm{k}, s} & = -(1 - \rho_s) \left( |\bm{k}|^2 - \frac{1}{12} {k_x}^4 - \frac{1}{12} {k_y}^4 \right) \nu_{\bm{k}, s} - \varepsilon_s (1 - \rho_s) \left[ i k_x w_{x, \bm{k}, s} - i k_y w_{y, \bm{k}, s} + \frac{1}{4} {k_x}^2 (\rho_{\bm{k}, s} + \nu_{\bm{k}, s}) - \frac{1}{4} {k_y}^2 (\rho_{\bm{k}, s} - \nu_{\bm{k}, s}) \right] \nonumber \\
    & \quad - \varepsilon_s (1 - \rho_s) \left[ -\frac{i}{6} {k_x}^3 w_{x, \bm{k}, s} + \frac{i}{6} {k_y}^3 w_{y, \bm{k}, s} - \frac{1}{48} {k_x}^4 (\rho_{\bm{k}, s} + \nu_{\bm{k}, s}) + \frac{1}{48} {k_y}^4 (\rho_{\bm{k}, s} - \nu_{\bm{k}, s}) \right]. \label{eq_mean_field_1site_linearized_nu}
\end{align}

To find the linear instability condition for $\bm{k} \to \bm{0}$, which corresponds to macroscopic phase separation, we apply the adiabatic approximation~\cite{Speck2014, Speck2015} to the equations for fast variables $\bm{w}_{\bm{k}, s}$ [Eqs.~\eqref{eq_mean_field_1site_linearized_wx} and \eqref{eq_mean_field_1site_linearized_wy}] and $\nu_{\bm{k}, s}$ [Eq.~\eqref{eq_mean_field_1site_linearized_nu}], which leads to
\begin{align}
    w_{x, \bm{k}, s} & = \frac{i}{4 \gamma} \varepsilon_s (2 \rho_s - 1) \left( k_x - \frac{1}{6} {k_x}^3 \right) \rho_{\bm{k}, s} - \frac{i}{8 \gamma^2} \varepsilon_s \left( 1 + \frac{1}{4} \varepsilon_{\bar{s}} \right) (1 - \rho_s) (2 \rho_s - 1) k_x |\bm{k}|^2 \rho_{\bm{k}, s} \nonumber \\
    & \quad - \frac{i}{16 \gamma^2} {\varepsilon_s}^2 (1 - \rho_s) (2 \rho_s - 1) {k_x}^3 \rho_{\bm{k}, s} - \frac{i}{4 \gamma} \varepsilon_s (1 - \rho_s) k_x \nu_{\bm{k}, s} + O (|\bm{k}|^5 \rho_{\bm{k}, s}, |\bm{k}|^3 \nu_{\bm{k}, s}), \\
    w_{y, \bm{k}, s} & = \frac{i}{4 \gamma} \varepsilon_s (2 \rho_s - 1) \left( k_y - \frac{1}{6} {k_y}^3 \right) \rho_{\bm{k}, s} - \frac{i}{8 \gamma^2} \varepsilon_s \left( 1 + \frac{1}{4} \varepsilon_{\bar{s}} \right) (1 - \rho_s) (2 \rho_s - 1) k_y |\bm{k}|^2 \rho_{\bm{k}, s} \nonumber \\
    & \quad - \frac{i}{16 \gamma^2} {\varepsilon_s}^2 (1 - \rho_s) (2 \rho_s - 1) {k_y}^3 \rho_{\bm{k}, s} + \frac{i}{4 \gamma} \varepsilon_s (1 - \rho_s) k_y \nu_{\bm{k}, s} + O (|\bm{k}|^5 \rho_{\bm{k}, s}, |\bm{k}|^3 \nu_{\bm{k}, s}), \\
    \nu_{\bm{k}, s} & = -\frac{1}{16 \gamma} \varepsilon_s ({k_x}^2 - {k_y}^2) \rho_{\bm{k}, s} - \frac{i}{4 \gamma} \varepsilon_s (1 - \rho_s) (k_x w_{x, \bm{k}, s} - k_y w_{y, \bm{k}, s}) + O(|\bm{k}|^4 \rho_{\bm{k}, s}, |\bm{k}|^2 \bm{k} \cdot \bm{w}_{\bm{k}, s}).
\end{align}
By substituting these into Eq.~\eqref{eq_mean_field_1site_linearized_rho}, we obtain
\begin{equation}
    \partial_t \rho_{\bm{k}, \mathrm{solute}} = r_{\bm{k}}^\mathrm{MF1} \rho_{\bm{k}, \mathrm{solute}}
\end{equation}
with
\begin{equation}
    r_{\bm{k}}^\mathrm{MF1} = c_2^\mathrm{MF1} |\bm{k}|^2 + c_{4+}^\mathrm{MF1} |\bm{k}|^4 + c_{4-}^\mathrm{MF1} ({k_x}^2 - {k_y}^2)^2 + O(|\bm{k}|^6).
\end{equation}
Here, the coefficients are
\begin{align}
    c_2^\mathrm{MF1} & = -\left( 1 + \frac{\varepsilon_\mathrm{solute} + \varepsilon_\mathrm{solvent}}{4} \right) + \frac{{\varepsilon_\mathrm{solute}}^2 (1 - \rho) (2 \rho - 1)}{4 \gamma} - \frac{{\varepsilon_\mathrm{solvent}}^2 \rho (2 \rho - 1)}{4 \gamma}, \\
    c_{4+}^\mathrm{MF1} & = \frac{1}{24} \left( 1 + \frac{\varepsilon_\mathrm{solute} + \varepsilon_\mathrm{solvent}}{4} \right) - \frac{{\varepsilon_\mathrm{solute}}^2 (1 - \rho) (2 \rho - 1)}{24 \gamma} + \frac{{\varepsilon_\mathrm{solvent}}^2 \rho (2 \rho - 1)}{24 \gamma} \nonumber \\
    & \quad - \frac{{\varepsilon_\mathrm{solute}}^2}{8 \gamma^2} \left( 1 + \frac{\varepsilon_\mathrm{solute} + \varepsilon_\mathrm{solvent}}{4} \right) (1 - \rho)^2 (2 \rho - 1) + \frac{{\varepsilon_\mathrm{solvent}}^2}{8 \gamma^2} \left( 1 + \frac{\varepsilon_\mathrm{solute} + \varepsilon_\mathrm{solvent}}{4} \right) \rho^2 (2 \rho - 1), \\
    c_{4-}^\mathrm{MF1} & = \frac{1}{24} \left( 1 + \frac{\varepsilon_\mathrm{solute} + \varepsilon_\mathrm{solvent}}{4} \right) - \frac{{\varepsilon_\mathrm{solute}}^2 (1 - \rho) (16 \rho - 11)}{192 \gamma} + \frac{{\varepsilon_\mathrm{solvent}}^2 \rho (16 \rho - 5)}{192 \gamma} \nonumber \\
    & \quad - \frac{{\varepsilon_\mathrm{solute}}^3 (1 - \rho)^2 (3 \rho - 2)}{32 \gamma^2} + \frac{{\varepsilon_\mathrm{solvent}}^3 \rho^2 (3 \rho - 1)}{32 \gamma^2} - \frac{{\varepsilon_\mathrm{solute}}^4 (1 - \rho)^3 (2 \rho - 1)}{64 \gamma^3} + \frac{{\varepsilon_\mathrm{solvent}}^4 \rho^3 (2 \rho - 1)}{64 \gamma^3}.
\end{align}
We numerically confirmed that $c_{4-}^\mathrm{MF1} < 0$ near the Lifshitz point, which is determined by $c_2^\mathrm{MF1} = 0$ and $c_{4+}^\mathrm{MF1} = 0$.
This means that, near the Lifshitz point, the diagonal modes (i.e., $k_x = \pm k_y$) are most unstable, compared with other fluctuation modes with the same wavenumber $|\bm{k}|$.

\subsection{Explicit form of dynamical equation in four-site mean-field theory}
\label{app_explicit_form_of_dynamical_mean-field_equation}

We consider the mean-field dynamical equation for the four-site density $\lambda_{i, \underline{s}, \underline{p}} (t)$, where $\underline{s} = (s_1, s_2, s_3, s_4)$ and $\underline{p} = (p_1, p_2, p_3, p_4)$ represent the configuration for a square cluster $i$, which consists of four sites: $i_1 = i$, $i_2 = i + \hat{x}$, $i_3 = i + \hat{x} + \hat{y}$, and $i_4 = i + \hat{y}$.
Neglecting the correlations across square clusters [Fig.~\ref{fig_mean-field_schematic}(b)], we obtain Eq.~\eqref{eq_mean_field_4site}:
\begin{equation}
    \partial_t \lambda_{i, \underline{s}, \underline{p}} = g_{i, \underline{s}, \underline{p}}^{(p, w)} + g_{i, \underline{s}, \underline{p}}^{(a, w)} + g_{i, \underline{s}, \underline{p}}^{(p, n)} + g_{i, \underline{s}, \underline{p}}^{(a, n)} + g_{i, \underline{s}, \underline{p}}^{(r)}.
\end{equation}

The first term, $g_{i, \underline{s}, \underline{p}}^{(p, w)}$, is the contribution from the passive exchange within cluster $i$.
We can divide $g_{i, \underline{s}, \underline{p}}^{(p, w)}$ into four terms:
\begin{equation}
    g_{i, \underline{s}, \underline{p}}^{(p, w)} = g_{i, \underline{s}, \underline{p}}^{(p, w, 12)} + g_{i, \underline{s}, \underline{p}}^{(p, w, 23)} + g_{i, \underline{s}, \underline{p}}^{(p, w, 34)} + g_{i, \underline{s}, \underline{p}}^{(p, w, 41)},
\end{equation}
where superscript $b$ in $g_{i, \underline{s}, \underline{p}}^{(p, w, b)}$ suggests the bond at which the passive exchange occurs (e.g., $g_{i, \underline{s}, \underline{p}}^{(p, w, 23)}$ corresponds to the exchange of a particle at site $i_2 = i + \hat{x}$ and a particle at site $i_3 = i + \hat{x} + \hat{y}$).
Specifically,
\begin{align}
    g_{i, \underline{s}, \underline{p}}^{(p, w, 12)} & = \delta_{s_1, \bar{s}_2} (\lambda_{i, \underline{s}^{12},\underline{p}^{12}} - \lambda_{i, \underline{s}, \underline{p}}) \label{appeq_g_p_w_12} \\
    g_{i, \underline{s}, \underline{p}}^{(p, w, 23)} & = \delta_{s_2, \bar{s}_3} (\lambda_{i, \underline{s}^{23},\underline{p}^{23}} - \lambda_{i, \underline{s}, \underline{p}}) \\
    g_{i, \underline{s}, \underline{p}}^{(p, w, 34)} & = \delta_{s_3, \bar{s}_4} (\lambda_{i, \underline{s}^{34},\underline{p}^{34}} - \lambda_{i, \underline{s}, \underline{p}}) \\
    g_{i, \underline{s}, \underline{p}}^{(p, w, 41)} & = \delta_{s_4, \bar{s}_1} (\lambda_{i, \underline{s}^{41},\underline{p}^{41}} - \lambda_{i, \underline{s}, \underline{p}}), \label{appeq_g_p_w_41}
\end{align}
where $\bar{s}$ represents the species opposite to $s$ (e.g., $\overline{\mathrm{solute}} = \mathrm{solvent}$), and superscript $b$ in $\underline{s}^b$ or $\underline{p}^b$ suggests the interchanged indices [e.g., $\underline{s}^{23} = (s_1, s_3, s_2, s_4)$].
The first and second terms in each of Eqs.~\eqref{appeq_g_p_w_12}-\eqref{appeq_g_p_w_41} represent the gain and loss of state $(\underline{s}, \underline{p})$ at $(i_1, i_2, i_3, i_4)$, respectively.
Note that Eqs.~\eqref{appeq_g_p_w_12}-\eqref{appeq_g_p_w_41} are derived from the master equation with no approximation.

Similarly, $g_{i, \underline{s}, \underline{p}}^{(a, w)}$ is the contribution from the active exchange within cluster $i$, which is divided into four terms:
\begin{equation}
    g_{i, \underline{s}, \underline{p}}^{(a, w)} = g_{i, \underline{s}, \underline{p}}^{(a, w, 12)} + g_{i, \underline{s}, \underline{p}}^{(a, w, 23)} + g_{i, \underline{s}, \underline{p}}^{(a, w, 34)} + g_{i, \underline{s}, \underline{p}}^{(a, w, 41)}.
\end{equation}
Specifically,
\begin{align}
    g_{i, \underline{s}, \underline{p}}^{(a, w, 12)} & = \delta_{s_1, \bar{s}_2} [(\varepsilon_{s_1} \delta_{p_1, -\hat{x}} + \varepsilon_{s_2} \delta_{p_2, +\hat{x}}) \lambda_{i, \underline{s}^{12},\underline{p}^{12}} - (\varepsilon_{s_1} \delta_{p_1, +\hat{x}} + \varepsilon_{s_2} \delta_{p_2, -\hat{x}}) \lambda_{i, \underline{s}, \underline{p}}] \label{appeq_g_a_w_12} \\
    g_{i, \underline{s}, \underline{p}}^{(a, w, 23)} & = \delta_{s_2, \bar{s}_3} [(\varepsilon_{s_2} \delta_{p_2, -\hat{y}} + \varepsilon_{s_3} \delta_{p_3, +\hat{y}}) \lambda_{i, \underline{s}^{23},\underline{p}^{23}} - (\varepsilon_{s_2} \delta_{p_2, +\hat{y}} + \varepsilon_{s_3} \delta_{p_3, -\hat{y}}) \lambda_{i, \underline{s}, \underline{p}}] \\
    g_{i, \underline{s}, \underline{p}}^{(a, w, 34)} & = \delta_{s_3, \bar{s}_4} [(\varepsilon_{s_3} \delta_{p_3, +\hat{x}} + \varepsilon_{s_4} \delta_{p_4, -\hat{x}}) \lambda_{i, \underline{s}^{34},\underline{p}^{34}} - (\varepsilon_{s_3} \delta_{p_3, -\hat{x}} + \varepsilon_{s_4} \delta_{p_4, +\hat{x}}) \lambda_{i, \underline{s}, \underline{p}}] \\
    g_{i, \underline{s}, \underline{p}}^{(a, w, 41)} & = \delta_{s_4, \bar{s}_1} [(\varepsilon_{s_4} \delta_{p_4, +\hat{y}} + \varepsilon_{s_1} \delta_{p_1, -\hat{y}}) \lambda_{i, \underline{s}^{41},\underline{p}^{41}} - (\varepsilon_{s_4} \delta_{p_4, -\hat{y}} + \varepsilon_{s_1} \delta_{p_1, +\hat{y}}) \lambda_{i, \underline{s}, \underline{p}}] \label{appeq_g_a_w_41}
\end{align}
Note that Eqs.~\eqref{appeq_g_a_w_12}-\eqref{appeq_g_a_w_41} are derived from the master equation with no approximation in a similar way as Eqs.~\eqref{appeq_g_p_w_12}-\eqref{appeq_g_p_w_41}.

Next, $g_{i, \underline{s}, \underline{p}}^{(p, n)}$ corresponds to the passive exchange across cluster $i$ and neighboring clusters.
We can divide $g_{i, \underline{s}, \underline{p}}^{(p, n)}$ into eight terms:
\begin{equation}
    g_{i, \underline{s}, \underline{p}}^{(p, n)} = g_{i, \underline{s}, \underline{p}}^{(p, n, 14')} + g_{i, \underline{s}, \underline{p}}^{(p, n, 23')} + g_{i, \underline{s}, \underline{p}}^{(p, n, 21')} + g_{i, \underline{s}, \underline{p}}^{(p, n, 34')} + g_{i, \underline{s}, \underline{p}}^{(p, n, 32')} + g_{i, \underline{s}, \underline{p}}^{(p, n, 41')} + g_{i, \underline{s}, \underline{p}}^{(p, n, 43')} + g_{i, \underline{s}, \underline{p}}^{(p, n, 12')},
\end{equation}
where superscript $b$ in $g_{i, \underline{s}, \underline{p}}^{(p, n, b)}$ suggests the bond at which the passive exchange occurs, and ($^\prime$) suggests the site index within the neighboring cluster (e.g., $g_{i, \underline{s}, \underline{p}}^{(p, n, 14')}$ corresponds to the exchange of a particle at site $i_1 = i$ and a particle at site $i_4 - 2\hat{y} = i - \hat{y}$, and $g_{i, \underline{s}, \underline{p}}^{(p, n, 21')}$ corresponds to the exchange of a particle at site $i_2 = i + \hat{x}$ and a particle at site $i_1 + 2 \hat{x} = i + 2 \hat{x}$).
Specifically,
\begin{align}
    g_{i, \underline{s}, \underline{p}}^{(p, n, 14')} & = \sum_{\underline{s}', \underline{p}'} \delta_{s_1, \bar{s}'_4} (\lambda_{i, \underline{s}^{1 \to 4'}, \underline{p}^{1 \to 4'}} \lambda_{i - 2 \hat{y}, \underline{s}'^{4' \to 1}, \underline{p}'^{4' \to 1}} - \lambda_{i, \underline{s}, \underline{p}} \lambda_{i - 2 \hat{y}, \underline{s}', \underline{p}'}) \label{appeq_g_p_n_14p} \\
    g_{i, \underline{s}, \underline{p}}^{(p, n, 23')} & = \sum_{\underline{s}', \underline{p}'} \delta_{s_2, \bar{s}'_3} (\lambda_{i, \underline{s}^{2 \to 3'}, \underline{p}^{2 \to 3'}} \lambda_{i - 2 \hat{y}, \underline{s}'^{3' \to 2}, \underline{p}'^{3' \to 2}} - \lambda_{i, \underline{s}, \underline{p}} \lambda_{i - 2 \hat{y}, \underline{s}', \underline{p}'}) \\
    g_{i, \underline{s}, \underline{p}}^{(p, n, 21')} & = \sum_{\underline{s}', \underline{p}'} \delta_{s_2, \bar{s}'_1} (\lambda_{i, \underline{s}^{2 \to 1'}, \underline{p}^{2 \to 1'}} \lambda_{i + 2 \hat{x}, \underline{s}'^{1' \to 2}, \underline{p}'^{1' \to 2}} - \lambda_{i, \underline{s}, \underline{p}} \lambda_{i + 2 \hat{x}, \underline{s}', \underline{p}'}) \\
    g_{i, \underline{s}, \underline{p}}^{(p, n, 34')} & = \sum_{\underline{s}', \underline{p}'} \delta_{s_3, \bar{s}'_4} (\lambda_{i, \underline{s}^{3 \to 4'}, \underline{p}^{3 \to 4'}} \lambda_{i + 2 \hat{x}, \underline{s}'^{4' \to 3}, \underline{p}'^{4' \to 3}} - \lambda_{i, \underline{s}, \underline{p}} \lambda_{i + 2 \hat{x}, \underline{s}', \underline{p}'}) \\
    g_{i, \underline{s}, \underline{p}}^{(p, n, 32')} & = \sum_{\underline{s}', \underline{p}'} \delta_{s_3, \bar{s}'_2} (\lambda_{i, \underline{s}^{3 \to 2'}, \underline{p}^{3 \to 2'}} \lambda_{i + 2 \hat{y}, \underline{s}'^{2' \to 3}, \underline{p}'^{2' \to 3}} - \lambda_{i, \underline{s}, \underline{p}} \lambda_{i + 2 \hat{y}, \underline{s}', \underline{p}'}) \\
    g_{i, \underline{s}, \underline{p}}^{(p, n, 41')} & = \sum_{\underline{s}', \underline{p}'} \delta_{s_4, \bar{s}'_1} (\lambda_{i, \underline{s}^{4 \to 1'}, \underline{p}^{4 \to 1'}} \lambda_{i + 2 \hat{y}, \underline{s}'^{1' \to 4}, \underline{p}'^{1' \to 4}} - \lambda_{i, \underline{s}, \underline{p}} \lambda_{i + 2 \hat{y}, \underline{s}', \underline{p}'}) \\
    g_{i, \underline{s}, \underline{p}}^{(p, n, 43')} & = \sum_{\underline{s}', \underline{p}'} \delta_{s_4, \bar{s}'_3} (\lambda_{i, \underline{s}^{4 \to 3'}, \underline{p}^{4 \to 3'}} \lambda_{i - 2 \hat{x}, \underline{s}'^{3' \to 4}, \underline{p}'^{3' \to 4}} - \lambda_{i, \underline{s}, \underline{p}} \lambda_{i - 2 \hat{x}, \underline{s}', \underline{p}'}) \\
    g_{i, \underline{s}, \underline{p}}^{(p, n, 12')} & = \sum_{\underline{s}', \underline{p}'} \delta_{s_1, \bar{s}'_2} (\lambda_{i, \underline{s}^{1 \to 2'}, \underline{p}^{1 \to 2'}} \lambda_{i - 2 \hat{x}, \underline{s}'^{2' \to 1}, \underline{p}'^{2' \to 1}} - \lambda_{i, \underline{s}, \underline{p}} \lambda_{i - 2 \hat{x}, \underline{s}', \underline{p}'}), \label{appeq_g_p_n_12p}
\end{align}
where superscript $c$ in $\underline{s}^c$ or $\underline{p}^c$ suggests the changed index [e.g., $\underline{s}^{1 \to 4'} = (s'_4, s_2, s_3, s_4)$ and $\underline{s}'^{3' \to 2} = (s'_1, s'_2, s_2, s'_4)$].
Note that the nonlinearity in Eqs.~\eqref{appeq_g_p_n_14p}-\eqref{appeq_g_p_n_12p} is derived from the master equation by neglecting the correlation between different clusters.

Similarly, $g_{i, \underline{s}, \underline{p}}^{(a, n)}$ represents the contribution from the active exchange across cluster $i$ and neighboring clusters, which is divided into eight terms:
\begin{equation}
    g_{i, \underline{s}, \underline{p}}^{(a, n)} = g_{i, \underline{s}, \underline{p}}^{(a, n, 14')} + g_{i, \underline{s}, \underline{p}}^{(a, n, 23')} + g_{i, \underline{s}, \underline{p}}^{(a, n, 21')} + g_{i, \underline{s}, \underline{p}}^{(a, n, 34')} + g_{i, \underline{s}, \underline{p}}^{(a, n, 32')} + g_{i, \underline{s}, \underline{p}}^{(a, n, 41')} + g_{i, \underline{s}, \underline{p}}^{(a, n, 43')} + g_{i, \underline{s}, \underline{p}}^{(a, n, 12')}.
\end{equation}
Specifically,
\begin{align}
    g_{i, \underline{s}, \underline{p}}^{(a, n, 14')} & = \sum_{\underline{s}', \underline{p}'} \delta_{s_1, \bar{s}'_4} [(\varepsilon_{s_1} \delta_{p_1, +\hat{y}} + \varepsilon_{s'_4} \delta_{p'_4, -\hat{y}}) \lambda_{i, \underline{s}^{1 \to 4'}, \underline{p}^{1 \to 4'}} \lambda_{i - 2 \hat{y}, \underline{s}'^{4' \to 1}, \underline{p}'^{4' \to 1}} - (\varepsilon_{s_1} \delta_{p_1, -\hat{y}} + \varepsilon_{s'_4} \delta_{p'_4, +\hat{y}}) \lambda_{i, \underline{s}, \underline{p}} \lambda_{i - 2 \hat{y}, \underline{s}', \underline{p}'}] \label{appeq_g_a_n_14p} \\
    g_{i, \underline{s}, \underline{p}}^{(a, n, 23')} & = \sum_{\underline{s}', \underline{p}'} \delta_{s_2, \bar{s}'_3} [(\varepsilon_{s_2} \delta_{p_2, +\hat{y}} + \varepsilon_{s'_3} \delta_{p'_3, -\hat{y}}) \lambda_{i, \underline{s}^{2 \to 3'}, \underline{p}^{2 \to 3'}} \lambda_{i - 2 \hat{y}, \underline{s}'^{3' \to 2}, \underline{p}'^{3' \to 2}} - (\varepsilon_{s_2} \delta_{p_2, -\hat{y}} + \varepsilon_{s'_3} \delta_{p'_3, +\hat{y}}) \lambda_{i, \underline{s}, \underline{p}} \lambda_{i - 2 \hat{y}, \underline{s}', \underline{p}'}] \\
    g_{i, \underline{s}, \underline{p}}^{(a, n, 21')} & = \sum_{\underline{s}', \underline{p}'} \delta_{s_2, \bar{s}'_1} [(\varepsilon_{s_2} \delta_{p_2, -\hat{x}} + \varepsilon_{s'_1} \delta_{p'_1, +\hat{x}}) \lambda_{i, \underline{s}^{2 \to 1'}, \underline{p}^{2 \to 1'}} \lambda_{i + 2 \hat{x}, \underline{s}'^{1' \to 2}, \underline{p}'^{1' \to 2}} - (\varepsilon_{s_2} \delta_{p_2, +\hat{x}} + \varepsilon_{s'_1} \delta_{p'_1, -\hat{x}}) \lambda_{i, \underline{s}, \underline{p}} \lambda_{i + 2 \hat{x}, \underline{s}', \underline{p}'}] \\
    g_{i, \underline{s}, \underline{p}}^{(a, n, 34')} & = \sum_{\underline{s}', \underline{p}'} \delta_{s_3, \bar{s}'_4} [(\varepsilon_{s_3} \delta_{p_3, -\hat{x}} + \varepsilon_{s'_4} \delta_{p'_4, +\hat{x}}) \lambda_{i, \underline{s}^{3 \to 4'}, \underline{p}^{3 \to 4'}} \lambda_{i + 2 \hat{x}, \underline{s}'^{4' \to 3}, \underline{p}'^{4' \to 3}} - (\varepsilon_{s_3} \delta_{p_3, +\hat{x}} + \varepsilon_{s'_4} \delta_{p'_4, -\hat{x}}) \lambda_{i, \underline{s}, \underline{p}} \lambda_{i + 2 \hat{x}, \underline{s}', \underline{p}'}] \\
    g_{i, \underline{s}, \underline{p}}^{(a, n, 32')} & = \sum_{\underline{s}', \underline{p}'} \delta_{s_3, \bar{s}'_2} [(\varepsilon_{s_3} \delta_{p_3, -\hat{y}} + \varepsilon_{s'_2} \delta_{p'_2, +\hat{y}}) \lambda_{i, \underline{s}^{3 \to 2'}, \underline{p}^{3 \to 2'}} \lambda_{i + 2 \hat{y}, \underline{s}'^{2' \to 3}, \underline{p}'^{2' \to 3}} - (\varepsilon_{s_3} \delta_{p_3, +\hat{y}} + \varepsilon_{s'_2} \delta_{p'_2, -\hat{y}}) \lambda_{i, \underline{s}, \underline{p}} \lambda_{i + 2 \hat{y}, \underline{s}', \underline{p}'}] \\
    g_{i, \underline{s}, \underline{p}}^{(a, n, 41')} & = \sum_{\underline{s}', \underline{p}'} \delta_{s_4, \bar{s}'_1} [(\varepsilon_{s_4} \delta_{p_4, -\hat{y}} + \varepsilon_{s'_1} \delta_{p'_1, +\hat{y}}) \lambda_{i, \underline{s}^{4 \to 1'}, \underline{p}^{4 \to 1'}} \lambda_{i + 2 \hat{y}, \underline{s}'^{1' \to 4}, \underline{p}'^{1' \to 4}} - (\varepsilon_{s_4} \delta_{p_4, +\hat{y}} + \varepsilon_{s'_1} \delta_{p'_1, -\hat{y}}) \lambda_{i, \underline{s}, \underline{p}} \lambda_{i + 2 \hat{y}, \underline{s}', \underline{p}'}] \\
    g_{i, \underline{s}, \underline{p}}^{(a, n, 43')} & = \sum_{\underline{s}', \underline{p}'} \delta_{s_4, \bar{s}'_3} [(\varepsilon_{s_4} \delta_{p_4, +\hat{x}} + \varepsilon_{s'_3} \delta_{p'_3, -\hat{x}}) \lambda_{i, \underline{s}^{4 \to 3'}, \underline{p}^{4 \to 3'}} \lambda_{i - 2 \hat{x}, \underline{s}'^{3' \to 4}, \underline{p}'^{3' \to 4}} - (\varepsilon_{s_4} \delta_{p_4, -\hat{x}} + \varepsilon_{s'_3} \delta_{p'_3, +\hat{x}}) \lambda_{i, \underline{s}, \underline{p}} \lambda_{i - 2 \hat{x}, \underline{s}', \underline{p}'}] \\
    g_{i, \underline{s}, \underline{p}}^{(a, n, 12')} & = \sum_{\underline{s}', \underline{p}'} \delta_{s_1, \bar{s}'_2} [(\varepsilon_{s_1} \delta_{p_1, +\hat{x}} + \varepsilon_{s'_2} \delta_{p'_2, -\hat{x}}) \lambda_{i, \underline{s}^{1 \to 2'}, \underline{p}^{1 \to 2'}} \lambda_{i - 2 \hat{x}, \underline{s}'^{2' \to 1}, \underline{p}'^{2' \to 1}} - (\varepsilon_{s_1} \delta_{p_1, -\hat{x}} + \varepsilon_{s'_2} \delta_{p'_2, +\hat{x}}) \lambda_{i, \underline{s}, \underline{p}} \lambda_{i - 2 \hat{x}, \underline{s}', \underline{p}'}]. \label{appeq_g_a_n_12p}
\end{align}
Note that the nonlinearity in Eqs.~\eqref{appeq_g_a_n_14p}-\eqref{appeq_g_a_n_12p} is derived from the master equation by neglecting the correlation between different clusters in a similar way as Eqs.~\eqref{appeq_g_p_n_14p}-\eqref{appeq_g_p_n_12p}.

Lastly, $g_{i, \underline{s}, \underline{p}}^{(r)}$ represents the rotation of polarity within cluster $i$:
\begin{equation}
    g_{i, \underline{s}, \underline{p}}^{(r)} = \sum_{n = 1}^4 \gamma (\lambda_{i, \underline{s}, \underline{p}^{n:R}} + \lambda_{i, \underline{s}, \underline{p}^{n:R^{-1}}} - 2 \lambda_{i, \underline{s}, \underline{p}}).
\end{equation}
Here, superscript $r$ in $\underline{p}^r$ suggests the index and direction for rotated polarity [e.g., $\underline{p}^{2:R} = (p_1, R p_2, p_3, p_4)$ and $\underline{p}^{3:R^{-1}} = (p_1, p_2, R^{-1} p_3, p_4)$], where $R$ and $R^{-1}$ represent counterclockwise and clockwise rotations by $90^\circ$, respectively [e.g., $R (+\hat{y}) = -\hat{x}$].

\end{widetext}

%

\end{document}